\documentclass[aps,prl,reprint,superscriptaddress]{revtex4-2}

\usepackage{graphicx}
\usepackage{amsmath}
\usepackage{amssymb}
\usepackage{float}
\usepackage[colorlinks,urlcolor=blue]{hyperref}
\usepackage[capitalise]{cleveref}
\usepackage{nicefrac}
\usepackage{xcolor}
\usepackage{bigints}
\usepackage[normalem]{ulem}
\usepackage{multirow}

\begin{document}

\title{Full Classification of Transport on an Equilibrated 5/2 Edge via Shot Noise}
	
\author{Sourav Manna}
\email{sourav.manna@weizmann.ac.il}
\affiliation{Department of Condensed Matter Physics, Weizmann Institute of Science, Rehovot 7610001, Israel}
\affiliation{Raymond and Beverly Sackler School of Physics and Astronomy, Tel-Aviv University, Tel Aviv 6997801, Israel}	
	
\author{Ankur Das}
\email{ankur.das@weizmann.ac.il}
\affiliation{Department of Condensed Matter Physics, Weizmann Institute of Science, Rehovot 7610001, Israel}

\author{Moshe Goldstein}
\affiliation{Raymond and Beverly Sackler School of Physics and Astronomy, Tel-Aviv University, Tel Aviv 6997801, Israel}
	
\author{Yuval Gefen}
\affiliation{Department of Condensed Matter Physics, Weizmann Institute of Science, Rehovot 7610001, Israel}

\begin{abstract}
The nature of the bulk topological order of the 5/2 non-Abelian fractional quantum Hall state and the steady-state of its edge are long-studied questions. The most promising non-Abelian model bulk states are the Pfaffian (Pf), anti-Pffafian (APf), and particle-hole symmetric Pfaffian (PHPf). 
Here, we propose to employ a set of dc current-current correlations \emph{(electrical shot noise)} 
in order to distinguish among the Pf, APf, and PHPf candidate states, as well as 
to determine their edge thermal
equilibration regimes: full vs.\ partial. 
Using other tools, measurements of GaAs platforms have already indicated consistency with the PHPf state. 
Our protocol, realizable with available experimental tools, is
based on fully electrical measurements.
\end{abstract}	
\maketitle

\textit{Introduction.---} 
Manifestations of  non-Abelian  braiding statistics
\cite{Moore-Read1} rely on foundational facets of  strongly-correlated many-body platforms, and may pave 
the way towards establishing tools for topological quantum computation \cite{arxiv.2210.10255}.
 A potential
host of such quasiparticles is the $5/2$ fractional quantum
Hall state \cite{PhysRevLett.59.1776,Kun2022}.
 As far as the nature of the bulk
state is concerned, there has been a number of proposals
which lead to the same electrical Hall conductance
but differ in their edge structure and the two-terminal
thermal conductance. The three most prominent non-Abelian model states are the Pfaffian (Pf) \cite{Moore-Read1}, its hole-conjugate
version --- the anti-Pfaffian (APf) \cite{PhysRevLett.99.236807, PhysRevLett.99.236806},
and the
particle-hole symmetric Pfaffian (PHPf) \cite{PhysRevX.3.041016, PhysRevX.5.031027, PhysRevLett.117.096802, PhysRevB.98.115107, PhysRevB.98.081107, PhysRevB.104.L081407}. 
Experimental measurements of two-terminal thermal conductance \cite{Banerjee2018} can be explained invoking either the APf or the PHPf state.
More recent experimental measurements of noise and thermal conductance are only consistent with the PHPf state \cite{Dutta2022,Dutta2022_Iso}.
This is in contradiction with numerical evidence 
\cite{PhysRevLett.80.1505, PhysRevLett.84.4685, PhysRevLett.101.016807, PhysRevLett.105.096802, PhysRevLett.104.076803, PhysRevLett.106.116801, PhysRevX.5.021004,PhysRevLett.100.166803, PhysRevB.79.115322}
supporting either the Pf or APf state.
Other candidates include the Abelian $331$, anti-$331$, $113$
and K$=8$ states \cite{Halperin1, PhysRevB.90.161306, PhysRevB.88.085317} and non-Abelian
$\text{SU}(2)_2$ and $\text{anti-}\text{SU}(2)_2$ states \cite{PhysRevLett.66.802, PhysRevB.40.8079}, which are not consistent with experimental
measurements \cite{Banerjee2018,Dutta2022,Dutta2022_Iso}.

Another important classification concerns the edge of such states. One may characterize
the
edge modes according to their degree of equilibration, which
varies with the length of the edge. Previous studies \cite{Banerjee2018, Dutta2022_Iso} have demonstrated that the degree of thermal equilibration,
reached among the chiral modes, can be modified by changing the length of the edge/interface \cite{Dutta2022, Dutta2022_Iso}.
Major differences in transport features may arise, depending on whether full or partial thermal equilibration is approached. Thus, 
characterization of thermal equilibration regimes is an
outstanding problem.
It has also been shown that the thermal equilibration length can be parametrically longer than the charge equilibration length
\cite{PhysRevLett.126.216803, Melcer2022, Kumar2022, Srivastav2022}.
It is then reasonable to assume full charge equilibration regardless the degree of thermal equilibration.

\begin{table}[!t]
\centering
	\includegraphics[width=\columnwidth,trim=0 40 0 10]{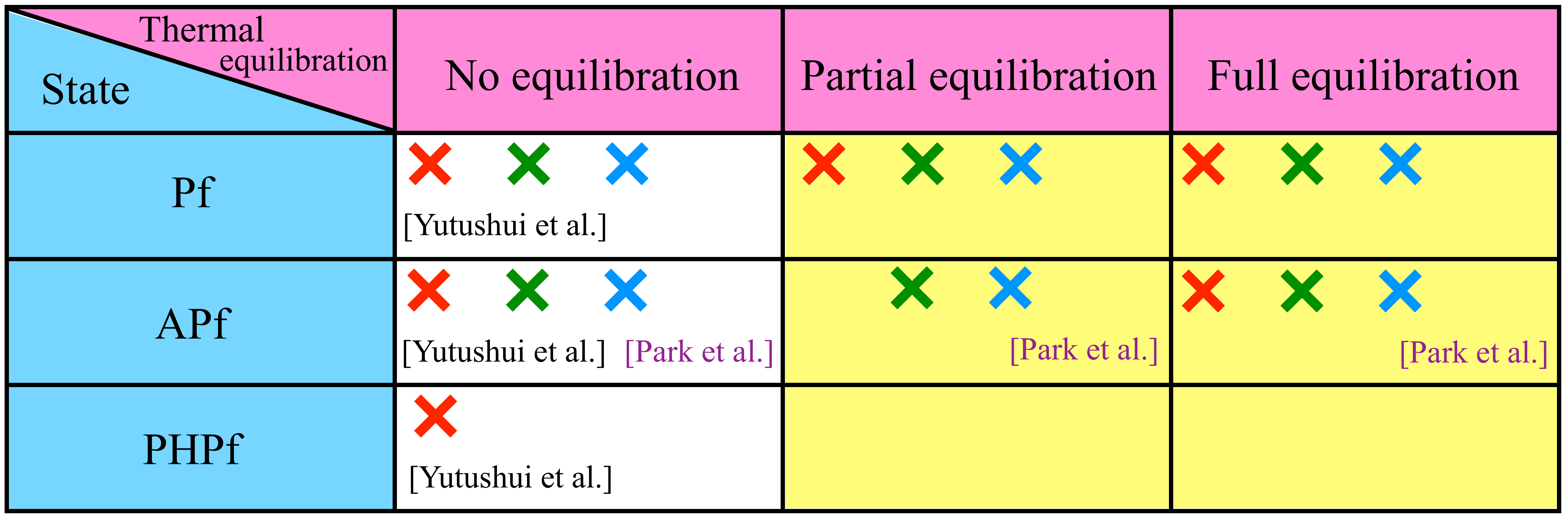}
	\caption{Current experimental and theoretical status for classification of the non-Abelian $5/2$ bulk states and their edge thermal equilibration regimes. Crosses (red, green, blue) represent the experimental exclusion
	of corresponding scenarios. Red, green, blue crosses exclude scenarios via measurements of two-terminal thermal conductance \cite{Banerjee2018}, interface noise \cite{Dutta2022}, and interface thermal conductance \cite{Dutta2022_Iso}, respectively. Theoretical proposals by Park et al. \cite{PhysRevLett.125.157702} and Yutushui et al. \cite{PhysRevLett.128.016401} used thermal transport with electrical shot noise at the edge and electrical transport with no equilibration, respectively, to confirm corresponding scenarios. Our work can distinguish among the yellow
	highlighted scenarios, thus completing the table.}\label{StateEq}
\end{table}

Measurements of shot noise and thermal conductance \cite{Dutta2022,
Dutta2022_Iso} in a device made out of interfaces of the $5/2$ state and Abelian
integer states are only consistent with the PHPf state.
A detailed theoretical model for the experiments in Ref.\ \onlinecite{Dutta2022,
Dutta2022_Iso} was put forward in Ref.\ \onlinecite{Michael2022}.
Preceding to those experiments, a number
of theoretical studies \cite{PhysRevB.78.161304, PhysRevB.87.125130, PhysRevB.93.201303, PhysRevB.97.121406,PhysRevB.97.165124,PhysRevB.98.045112, PhysRevLett.121.026801, PhysRevB.98.167401,
PhysRevB.99.085309,PhysRevB.101.041302,PhysRevB.102.205104,   PhysRevLett.125.146802,
PhysRevResearch.2.043242,
PhysRevLett.124.126801, PhysRevLett.125.157702} 
analyzed the Pf and APf states
to explain the first thermal measurements 
on the $5/2$ edge \cite{Banerjee2018} along with some proposals \cite{PhysRevB.78.161304, PhysRevB.104.L121106, PhysRevB.87.125130, PhysRevLett.125.157702, PhysRevLett.128.016401} to distinguish the bulk state. In particular, Ref.\ \onlinecite{PhysRevB.104.L121106} proposed that the Pf family members can be distinguished using their chiral
gravitons, which does not rely on edge physics.
Notably, Ref. \onlinecite{PhysRevLett.125.157702} has addressed the challenges of identifying the underlying state taking into account different equilibration regimes. It was shown that 
electrical shot noise combined with thermal transport can uniquely point out the APf state, along with the degree of equilibration. Those works did not
take into account that charge and thermal equilibration
lengths can differ by order of magnitudes \cite{PhysRevLett.126.216803, Melcer2022, Kumar2022, Srivastav2022}.
They also fall short (c.f.\  \cref{StateEq})
of providing unique diagnostics of the bulk 
topological state (Pf, APf, and PHPf)
and, simultaneously, the thermal equilibration regime.
This is the challenge addressed in this work.

\begin{figure}[!t]
\includegraphics[width=\columnwidth]{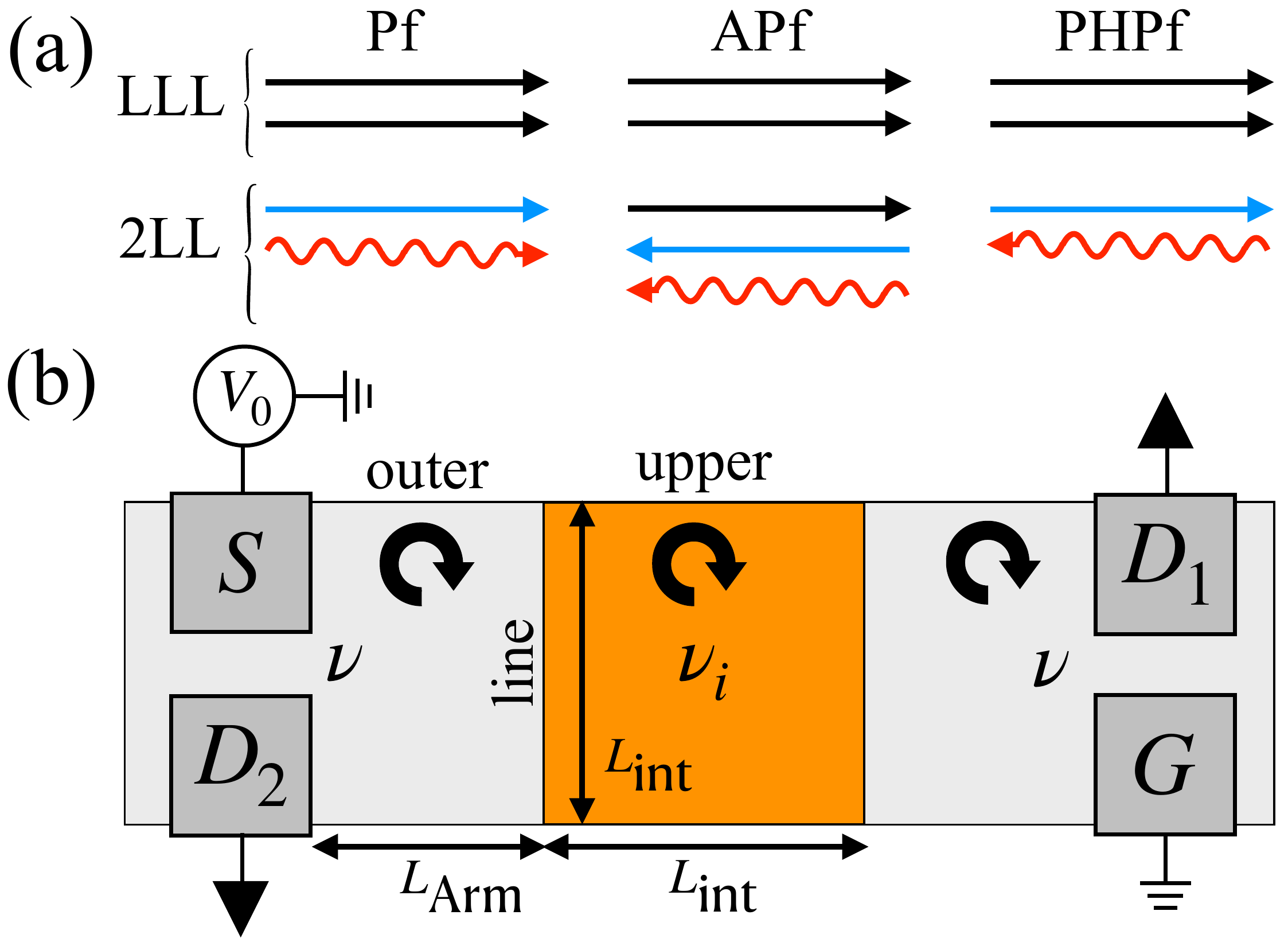}
\caption{(a) The edge structures of Pf, APf, and PHPf states: Integer, charge-$1/2$ boson \cite{Kun2022}, and charge-neutral Majorana modes are depicted as black, blue, and  wiggly red lines, respectively. Arrowheads show the chirality. The lowest Landau level (LLL) is same for all three states but the second Landau level (2LL) is different. (b) Our proposed device: A middle region of the Hall bar is depleted from filling $\nu$ to filling $\nu_i(<\nu)$ creating two interfaces (``line"). Specifically, we take
$\{\nu, \nu_i\}$ as $\{5/2,2\}$ or $\{5/2,7/3\}$ or $\{3,5/2\}$.
We call the boundary of the vacuum with $\nu$ the ``outer", and with $\nu_i$ the ``upper". We have four contacts, the source $S$ (biased by a dc voltage $V_0$), a ground contact $G$, and two drains $D_1, D_2$. Each circular arrow shows the chirality of charge propagation of the respective filling. The geometric lengths $L_{\text{Arm}}$ and $L_{\text{int}}$ are assumed to be of the same order of magnitude.}\label{fig:Device}
\end{figure}

Our work relies on the setup depicted in \cref{fig:Device}, where 
current-current correlations (CCC) are computed when the $5/2$ bulk state is interfaced with an
Abelian quantum Hall state. We calculate both dc auto- and cross-correlations (from now on we drop the ``dc" for simplicity).
Our results can discriminate between the full and partial  thermal equilibration regimes of the PHPf state. In addition, our protocol is unique in being able to distinguish among the Pf, APf, and PHPf states
along with their thermal equilibration regimes: full vs.\ partial.
Throughout the entire analysis, we assume that the charge equilibration
length (full equilibration has been observed over a distance 
$28\ \mu m$ at temperature 10 mK in GaAs \cite{Dutta2022} and $3\ \mu m$ at temperature 30 mK in
graphene \cite{PhysRevLett.126.216803}) is shorter than all the geometric lengths
and that there is 
no bulk-leakage
\cite{PhysRevLett.123.137701, Banerjee2018, PhysRevB.99.041302,PhysRevLett.125.157702}.
Our classification, depicted in \cref{Table}, relies on both a qualitative distinction between auto- and cross-correlations,
as well as quantitative differences. Our method applies for a uniform edge -- moving away 
from this assumption, e.g.\ having puddles of Pf and APf separated by domain walls \cite{PhysRevLett.121.026801, PhysRevB.98.045112}, would require technique beyond ours.
The many-puddle picture would facilitate 
heat leakage from the edge to the bulk \cite{MelcerBulk2023}. As long as the measured two-terminal heat conductance is half-integer 
quantized, it weakens the evidence for this non-uniform picture.

\textit{Device and thermal equilibration regimes.---}
Our device is made out of interfaces of the filling $\nu$ and filling $\nu_i(<\nu)$, c.f.\ \cref{fig:Device}.
We choose
$\{\nu, \nu_i\}$ as $\{5/2,2\}$ or $\{5/2,7/3\}$ or $\{3,5/2\}$ and we assume $\nu_i=7/3$ as an Abelian state \cite{PhysRevLett.107.036805, Banerjee2018, Dutta2022, Dutta2022_Iso}. The relevant lengths for this device are $(i)$ the internal characteristic lengths $l_{\text{eq}}^{\text{full}}$ and $l_{\text{eq}}^{\text{par}}$,
and $(ii)$ the geometric lengths $L_{\text{Arm}}$ and $L_{\text{int}}$ (cf.\ \cref{fig:Device}).
$l_{\text{eq}}^{\text{full}}$ denotes the full
thermal equilibration length (full equilibration has been observed over a distance 
$160\ \mu m$ at temperature 11 mK in GaAs \cite{Dutta2022_Iso}), over which the heat transfer among all modes is fully facilitated, i.e.,
$l_{\text{eq}}^{\text{full}} \ll (L_{\text{Arm}},L_{\text{int}}$) for full thermal
equilibration. $l_{\text{eq}}^{\text{par}}$ denotes the partial thermal
equilibration length, over which the heat transfer among modes of the same Landau level is fully facilitated but heat transfer among modes of different Landau Levels is negligible, i.e., $l_{\text{eq}}^{\text{par}}
\ll (L_{\text{Arm}},L_{\text{int}}) \ll l_{\text{eq}}^{\text{full}}$ for partial thermal equilibration. Equilibrated modes form a chiral hydrodynamic mode characterized by its electrical and thermal conductances. We refer to the direction parallel (anti-parallel) to the charge flow as downstream (upstream). We note that edge reconstruction cannot play any role in charge conductance due to 
full charge equilibration \cite{Srivastav2022, Kumar2022, Cohen2019,PhysRevLett.126.216803}. For full thermal equilibration, each mode
is equilibrated with other while for partial
thermal equilibration, each mode
is equilibrated with other only in a given Landau level.  As edge reconstruction takes place in each Landau level,
the effect of edge reconstruction is washed out in both the cases.

\begin{table}[!t]
  		\begin{tabular}{| c | c | c || c | c | c |}
			\hline

$\{\nu,\nu_i\}$    & State  & $|F^\text{Full}_{1,2,c}|$ & $F^\text{Partial}_1$   & $F^\text{Partial}_2$   & $|F^\text{Partial}_c|$ \\ \hline \hline
			\multirow{3}{*}{\{5/2,2\}}     & Pf & $\approx 0$ & $= 0$& $= 0$& $= 0$\\ \cline{2-6}
			& \textbf{APf} & $\approx 0.28$ & $= 0$     & $ \approx 0.12$ \cite{PhysRevLett.125.157702} & $= 0$\\ \cline{2-6}
			& PHPf & $\approx 0$ & $= 0$     & $= 0$   & $= 0$ \\ \hline\hline
			\multirow{3}{*}{\{5/2,7/3\}}      & Pf & $\approx 0$& $\approx 0$     & $\approx 0$   & $\approx 0$\\ \cline{2-6}
			& \textbf{APf} & $\approx 0.25$ & $\approx 0.5$     & $\approx 0.5$   & $\approx 0.27$\\ \cline{2-6}
			& \textbf{PHPf} & $\approx 0.11$ & $\approx 0.26$     & $\approx 0.26$   & $\approx 0.26$
			\\ 
			\hline \hline
			\multirow{3}{*}{\{3,5/2\}}      & \textbf{Pf} & $\approx 0.15$ & $\approx 0.19$     & $\approx 0.19$   & $\approx 0.19$\\ \cline{2-6}
			& \textbf{APf} & $\approx 0$ & $\approx 0.6$     & $\approx 0.6$   & $\approx 0.6$\\ \cline{2-6}
                & PHPf & $\approx 0$ & $\approx 0$     & $\approx 0$   & $\approx 0$
			\\\hline
		\end{tabular}
		\caption{Summary of our results: $F_1, F_2$ are the auto-correlation Fano factors for the drains $D_1, D_2$, respectively, and $F_c$ is the cross-correlation Fano factor. For full thermal equilibration (fourth column) we have $F_1=F_2=|F_c|$  $\equiv |F^\text{Full}_{1,2,c}|$. We use $\approx 0$ to indicate exponential suppression as a function of the geometric lengths.
	    }\label{Table}
\end{table}

\textit{Heat transport and current-current correlations.---}
The thermal conductance of a hydrodynamic mode is determined by the
difference of the thermal conductances of downstream and upstream modes ($\nu_Q$) involved in the 
equilibration process.
For $\nu_Q > 0$, $\nu_Q =0$, or $ \nu_Q < 0$, the nature of
heat transport in that hydrodynamic mode is, respectively, ballistic (B), diffusive (D), or antiballistic (AB) \cite{PhysRevLett.123.137701,PhysRevB.101.075308}. For the B, D, and AB heat transport 
we have, respectively, exponentially suppressed, algebraically decaying, and constant CCC as a function of the geometric length of the mode
\cite{PhysRevLett.123.137701, PhysRevB.101.075308}.

\begin{figure*}[!t]
	\includegraphics[width=\textwidth]{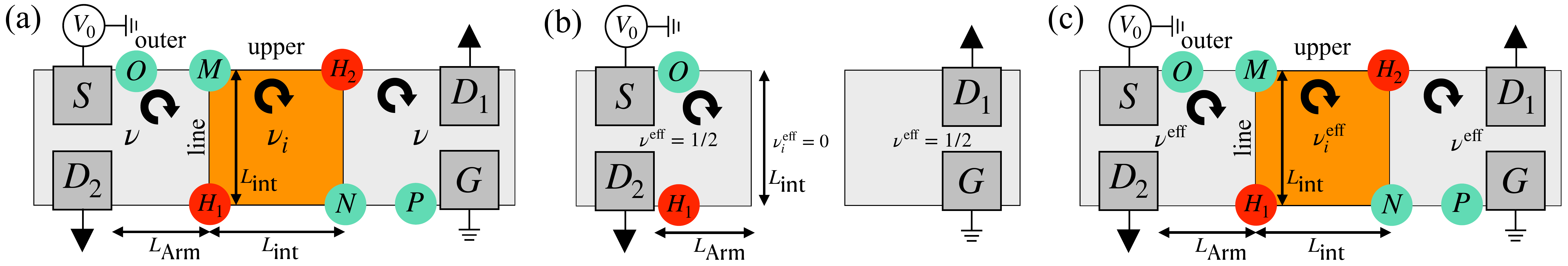}
	\caption{Noise generation in our proposed device (see \cref{fig:Device}) for full (a) and partial (b,c) thermal equilibration, where in the latter scenario the modes in the lowest Landau level localize in the ``line" segment. Each circular arrow indicates the chirality of charge propagation of the respective filling.  Voltage drops occur at the hot spots $H_1,H_2$ (red circles), resulting in the noise spots $M,N,O,P$ (green circles) \cite{PhysRevB.101.075308}. In (a) we have $\{\nu, \nu_i\}$ as $\{5/2,2\}$ or $\{5/2,7/3\}$ or $\{3,5/2\}$. In (b) we effectively have the filling $\{\nu^{\text{eff}}, \nu_i^{\text{eff}}\}=\{1/2,0\}$ for $\{\nu, \nu_i\}=\{5/2,2\}$. In (c) we effectively have the filling $\{\nu^{\text{eff}}, \nu_i^{\text{eff}}\}=\{1/2,1/3\}$ for $\{\nu, \nu_i\}=\{5/2,7/3\}$ and $\{\nu^{\text{eff}}, \nu_i^{\text{eff}}\}=\{1,1/2\}$ for $\{\nu, \nu_i\}=\{3,5/2\}$.}\label{Transport}
\end{figure*}

As contacts $S$ and $G$ in \cref{Transport} are at different potentials, there
are potential drops in the device which happen in the regions marked as hot spots, $H_1$, $H_2$, resulting in Joule heating \cite{PhysRevB.101.075308}.
Two possible hot spots can also exist near the drains $D_1, D_2$, though, heat generated there can not flow back to the middle
region due to their specific
configurations and hence can not contribute to the noise \cite{PhysRevB.101.075308}.
In addition,
four noise spots ($M,\ N,\ O,\ P$) are formed due to the creation of thermally excited particle-hole pairs
and their splitting into different drains $D_1, D_2$ (\cref{Transport}) \cite{PhysRevB.101.075308}. CCC are computed by collecting the contributions
from $M,N,O,P$.
Let us elaborate on how the charge-neutral Majorana mode $\psi$ contributes to CCC, as we know $\psi$ alone cannot participate in particle-hole pair splitting. 
However, we can define a creation operator $\psi e^{\mp 2 i \phi_\pm}$ ($\phi_\pm$ denotes downstream and upstream charge-$1/2$ boson, respectively) carrying an electron
charge. As $\psi$ always propagates alongside $\phi_\pm$, they together
can participate in particle-hole pair splitting.

The nature of heat transport in the ``outer", ``line" and ``upper" segments (cf.\ \cref{fig:Device} and \cref{Transport}) determine the CCC \cite{PhysRevB.101.075308, SMshort, SMLong, SMAD}. 
We define a quantity $\nu_Q^{\text{list}}$ as $ [\nu_Q\ \text{in ``outer"}, \nu_Q\ \text{in ``line"}, \nu_Q\ \text{in ``upper"}]$. In our analysis, $\nu_Q^{\text{list}}$ becomes either [B, B, B] or [B, AB, B] or
[AB, AB, B] or [B, B, AB].
 Hence, we have either exponentially suppressed or constant CCC as a function of the geometric lengths,
which we use to discriminate among the states and their thermal equilibration regimes. For the [B, B, B] case the generated heat at $H_1, H_2$ flows
downstream and reaches directly to the drains $D_1, D_2$. Only an exponentially suppressed, $\mathcal{O}(\exp{(-L_{\text{int}}}/l_{\text{eq}}))$ and $\mathcal{O}(\exp{(-L_{\text{Arm}}}/l_{\text{eq}}))$, amount of heat reaches $M, N$ and $O, P$ with $l_{\text{eq}} = l_{\text{eq}}^{\text{full}}$ and $l_{\text{eq}} =
l_{\text{eq}}^{\text{par}}$ for full and partial thermal equilibrations, respectively. Therefore 
an exponentially suppressed CCC is found.
For the [B, AB, B] and [B, B, AB]
cases the generated heat at $H_1, H_2$ flows upstream along the ``line" and ``upper" segments, respectively. Thus, $M, N$ provide constant contributions to the CCC, but the amount of heat that reaches $O, P$ is exponentially suppressed.
For the [AB, AB, B] case the generated heat at $H_1, H_2$ flows upstream along both the ``line" and ``outer"
segments. Thus both $M, N$ and $O, P$ provide constant contributions to CCC \cite{LLM}.
For partial thermal equilibration, we treat the lowest and second Landau levels dis-jointly while computing CCC.
Given impurity-facilitated inter-mode tunneling as well as inter-mode
interaction, equivalent (yet counter-propagating in $\nu$ and $\nu_i$) lowest Landau level modes become localized, decoupling them from the transport processes along the ``line".
Hence, the modes in the lowest Landau level do not contribute to CCC, while the modes in the second Landau level do contribute.

Let us denote by $I_1$ and $I_2$ the currents entering the drains $D_1$ and $D_2$, respectively. The corresponding current-current auto-correlations are defined as $\delta^2 I_1=\langle (I_1 - \langle I_1 \rangle)^2
\rangle$ in $D_1$ and $\delta^2 I_2=\langle (I_2 - \langle I_2 \rangle)^2 \rangle$ in $D_2$, while the cross-correlation is $\delta^2 I_c=\left\langle \left(I_1 - \langle I_1 \rangle\right)
\left(I_2 - \langle I_2 \rangle\right) \right\rangle$ \cite{averagedef}.
 The corresponding Fano factors are defined as $F_j = |\delta^2 I_j|/2e \langle I \rangle t(1-t)$, with $j \in \{1,2,c\}$, where $I$ is the source current and $t = \langle I_1 \rangle/\langle I \rangle$. The Fano factors are found to be \cite{SM} $F_1 = F_2 = F_O + F_P + F_M + F_N$ and $F_c = F'_O + F'_P - F_M - F_N$, where $F_\alpha$ is the contribution to the auto-correlations from the noise spot $\alpha \in \{M,N,O,P\}$ and $F'_\beta$ is the contribution to the cross-correlation from the noise spot $\beta \in \{O,P\}$.
We evaluate these contributions as
\begin{equation}\label{outernoise}
	\begin{split}
	&2e I t(1-t)(F_M+F_N)=2\frac{e^2}{h}(\nu-\nu_i)\frac{\nu_i}{\nu}k_{\text{B}}(T_M+T_N),\\&
	2e I t(1-t)(F_O+F_P)=\frac{1}{\nu^2}\left(\nu_i^2 S_O + (\nu-\nu_i)^2S_P\right),\\&
	2e I t(1-t)(F'_O+F'_P)=\frac{\nu_i(\nu-\nu_i)}{\nu^2}(S_O+S_P),
	\end{split}
\end{equation}
where $k_{\text{B}}$ is the Boltzmann constant, and we obtain the temperatures $T_M=T_N$ at $M, N$ by solving self-consistent equilibration equations and 
considering energy conservations \cite{PhysRevB.101.075308,SM}. We find the noise contributions $S_O=S_P$ by computing \cite{PhysRevB.101.075308,SM}
\begin{equation}\label{outernoise2}
	\begin{split}
		&S_O=S_P=\\&\frac{2e^2}{hl_{\text{eq}}^{\text{ch}}}\frac{\nu'\nu_{-}}{\nu_{+}} \bigintss_0^{L_{\text{Arm}}} \frac{e^{\frac{-2x}{l_{\text{eq}}^{\text{ch}}}}k_{\text{B}}\big[T_{+}(x)+T_{-}(x)\big]dx}{l_{\text{eq}}^{\text{ch}}\Big[1-\big(e^{\frac{-L_{\text{Arm}}}{l_{\text{eq}}^{\text{ch}}}}\frac{\nu_{-}}{\nu_{+}}\big)\Big]^2}
	\end{split}
\end{equation}
in the limit $l_{\text{eq}} \ll L_{\text{Arm}}$, where $\nu'=(\nu_{+}-\nu_{-})$.
Here $l_{\text{eq}}^{\text{ch}}$ is the charge equilibration length, and $T_{\pm}(x)$ (which depend on $l_{\text{eq}}$) are the temperature profiles in $\nu_{\pm}$, where $\nu_{+}$ and $\nu_{-}$ are the filling factors of the downstream
and the upstream mode in the ``outer" segment, respectively. 
$T_{\pm}(x)$
are calculated by solving the equations of
heat equilibration under the assumptions that no voltage drop occurs along the ``outer" segment and the lead contacts are at zero temperature. We note that in some cases we also have \emph{qualitative differences} between the auto-correlation and cross-correlation Fano
factors apart from their quantitative
differences (see \cref{Table}).

\textit{The case $\{\nu,\nu_i\}=\{5/2,2\}$.---}

(a) Full thermal equilibration.--- For the Pf  and PHPf states we have $\nu_Q^{\text{list}}=[7/2, 3/2, 2]$ and $\nu_Q^{\text{list}}=[5/2, 1/2, 2]$, respectively. The nature of heat 
transport is [B, B, B] for both states resulting in exponentially suppressed CCC. For the APf state we have $\nu_Q^{\text{list}}=[3/2, -1/2, 2]$, thus the nature of heat transport is [B,
AB, B]. In this case $F_O+F_P, F'_O+F'_P$ are exponentially suppressed and $F_M+F_N$ provides a constant contribution, leading to $F_1=F_2=|F_c| \neq 0$ (\cref{Table} and \cite{SM}).

(b) Partial thermal equilibration.--- The modes in the second Landau level equilibrate and form a hydrodynamic mode connecting $S, D_2$ and $G, D_1$ individually. Since only $S$ is biased, CCC can only appear in $D_2$, hence $F_1=|F_c|=0$ always.
For the Pf and PHPf states we have $\nu_Q=3/2$ and $\nu_Q=1/2$ for the mode connecting $S, D_2$ leading to the ballistic heat transport and an exponentially suppressed CCC. For the APf state we have $\nu_Q=-1/2$ for the mode connecting $S, D_2$ leading to  antiballistic heat transport and a constant CCC ($F_2 \neq 0$) (\cref{Table} and \cite{PhysRevLett.125.157702}).

\textit{The case $\{\nu,\nu_i\}=\{5/2,7/3\}$.---}

(a) Full thermal equilibration.--- For the Pf state we have $\nu_Q^{\text{list}}=[7/2, 1/2, 3]$. The nature of heat transport is [B, B, B], thus CCC are exponentially 
supressed. For the APf and PHPf states we have $\nu_Q^{\text{list}}=[3/2, -3/2, 3]$ and 
$\nu_Q^{\text{list}}=[5/2,-1/2, 3]$, respectively. The nature of heat transport is
[B, AB, B] for both the APf and PHPf states. In these cases
$F_O+F_P, F'_O+F'_P$ are exponentially suppressed and $F_M+F_N$ provides a constant contribution to the CCC, leading to $F_1=F_2=|F_c|\neq 0$ (\cref{Table} and \onlinecite{SM}).

(b) Partial thermal equilibration.--- 
For the Pf state we have $\nu_Q^{\text{list}}=[3/2, 1/2, 1]$.
The nature of heat transport is [B, B, B], thus CCC is exponentially 
supressed. For the APf state we have $\nu_Q^{\text{list}}=[-1/2, -3/2, 1]$. The nature of heat transport is then [AB, AB, B], hence $F_O+F_P, F'_O+F'_P, F_M+F_N$ all provide constant contributions, leading to $F_1=F_2 > |F_c| \neq 0$ (\cref{Table} and \cite{SM}). For the PHPf state we have $\nu_Q^{\text{list}}=[1/2, -1/2, 1]$. The nature of heat transport is therefore [B, AB, B]. In this case $F_O+F_P, F'_O+F'_P$ are exponentially suppressed and $F_M+F_N$ provides a constant contribution to the CCC, leading to $F_1=F_2=|F_c|\neq 0$ (\cref{Table} and \cite{SM}).

\textit{The case $\{\nu,\nu_i\}=\{3,5/2\}$.---}

(a) Full thermal equilibration.--- For the Pf state we have $\nu_Q^{\text{list}}=[3, -1/2, 7/2]$. The nature of heat transport is [B, AB, B]. In this case
$F_O+F_P, F'_O+F'_P$ are exponentially suppressed and $F_M+F_N$ provides a constant contribution to the CCC, leading to $F_1=F_2=|F_c|\neq 0$ (\cref{Table} and \onlinecite{SM}). For the APf and PHPf states we have $\nu_Q^{\text{list}}=[3, 3/2, 3/2]$ and 
$\nu_Q^{\text{list}}=[3, 1/2, 5/2]$. The nature of heat transport is
[B, B, B] for both states, thus CCC are exponentially 
supressed.

(b) Partial thermal equilibration.--- 
For the Pf and APf states we have $\nu_Q^{\text{list}}=[1, -1/2, 3/2]$ and $\nu_Q^{\text{list}}=[1, 3/2, -1/2]$, respectively, thus
the nature of heat transports are [B, AB, B] and [B, B, AB], respectively. In each case
$F_O+F_P, F'_O+F'_P$ are exponentially suppressed and $F_M+F_N$ provides a constant contribution to the CCC, leading to $F_1=F_2=|F_c|\neq 0$ (\cref{Table} and \onlinecite{SM}). For the PHPf state we have $\nu_Q^{\text{list}}=[1, 1/2, 1/2]$, leading to the nature of heat transport as
[B, B, B], thus CCC are exponentially 
supressed.

\textit{Protocol.---} 
Based on our results (\cref{Table}), we now present a sequential protocol to distinguish both the bulk Pf, APf, and PHPf states and the regimes of thermal equilibration: full vs.\ partial. First we choose $\{\nu,\nu_i\}=\{5/2,2\}$ and measure $F_1, F_2, F_c$. If $F_2
\neq 0$ then the state is APf. If in addition $F_1 = F_2 =
|F_c| \neq 0$ then we have full
thermal equilibration, while if $F_1 = |F_c| = 0$
then we have partial 
thermal equilibration.
If on the other hand $F_1=F_2=|F_c|=0$, then
the state can be either Pf or PHPf. Next we choose $\{\nu,\nu_i\}=\{5/2,7/3\}$. If $F_1=F_2=|F_c|\neq 0$ then the
state is PHPf. We can distinguish
between the full and partial thermal equilibration regimes
via the change in the Fano factors. 
If $F_1=F_2=|F_c| = 0$, then the
state is Pf. Thereafter, we choose $\{\nu,\nu_i\}=\{3,5/2\}$. The value of $F_1=F_2=|F_c| \neq 0$ 
distinguish between the full or partial
thermal equilibration regimes. We mention that, in principle, one can employ our method to distinguish among the $\text{SU}(2)_2$ family members (by suitably choosing \{$\nu, \nu_i$\}) \cite{PhysRevB.40.8079}, which are, however, inconsistent with experimental measurements \cite{Banerjee2018,Dutta2022,Dutta2022_Iso}.

\textit{Summary and outlook.---} We have considered a device comprising of the interface between the $5/2$ state with an Abelian state. We have
considered the Pf, APf, and PHPf states. We have studied
both the full and partial thermal equilibration regimes, where heat transfer among different Landau Levels is, respectively, fully facilitated or negligible. Throughout our analysis, the effective electro-chemical potential of the various edge modes is assumed to be fully equilibrated, leading to full charge equilibration \cite{PhysRevLett.126.216803, Melcer2022,
Kumar2022, Srivastav2022}. Our CCC-based protocol provides a new platform, different than other experimental platforms, capable of distinguishing among the Pf, APf, and PHPf states
along with their thermal equilibration regimes: full vs.\ partial (see \cref{Table}).
Despite of challenges and limitations to the experimental
implementation, we note that we just need two interfaces for making this device
and one can tune the filling in the middle region by changing the gate voltage.
Various experiments have already implemented devices based on interfaces both
in GaAs \cite{Biswas2022, Dutta2022, Dutta2022_Iso, Hashisaka2022, Cohen2019} and in graphene \cite{Chandan2018SSP, PhysRevB.98.155421, Paul2022}.

The application of the idea outlined here goes beyond the $5/2$ non-Abelian state in GaAs, for example classifying the $12/5$ state
\cite{PhysRevB.59.8084,PhysRevB.77.241306,PhysRevLett.93.176809,PhysRevLett.105.246808,PhysRevB.78.125323} 
and other Abelian states \cite{SMshort, SMLong, SMAD} in GaAs. They can also be extended to graphene \cite{Li2017, Huang2021, Ki2014, Zibrov2017} quantum Hall. Another interesting possibility is to
study the non-Abelian phase of $\alpha$-$\text{RuCl}_3$ Kitaev magnet \cite{Banerjee2016,Kasahara2018,Kitaev2006}.\\

\begin{acknowledgements}
We thank Alexander D.\ Mirlin, Bivas Dutta, Christian 
Sp\aa{}nsl\"att, Jinhong Park, Kun 
Yang, Matthew Foster, Moty Heiblum, Nandini Trivedi, and Subir Sachdev for useful
discussions. This research was supported in part by the International Centre for Theoretical Sciences (ICTS) for participating in the program - Condensed Matter meets Quantum Information (code: ICTS/COMQUI2023/9).
S.M.\ was supported by Weizmann Institute of Science, Israel Deans fellowship through Feinberg
Graduate School and the Raymond and Beverly Sackler Center for Computational Molecular and Material Science at Tel Aviv University. 
A.D.\ was supported by the German-Israeli Foundation Grant No. I-1505-303.10/2019, DFG
MI 658/10-2, DFG RO 2247/11-1, DFG EG 96/13-1,
and CRC 183 (project C01). A.D.\ also thanks the Israel planning and budgeting committee (PBC) and the
Weizmann Institute of Science, the Dean of Faculty fellowship, and the Koshland Foundation for financial support.
M.G.\ has been supported by the Israel Science Foundation (ISF) and the Directorate for Defense
Research and Development (DDR\&D) Grant No. 3427/21 and by the US-Israel Binational Science Foundation (BSF) Grant No. 2020072. 
Y.G.\ acknowledges support by the DFG Grant MI 658/10-2, by the German-Israeli Foundation Grant
I-1505-303.10/2019, by the Helmholtz International Fellow Award, by the DFG
Grant RO 2247/11-1, by CRC 183 (project C01), and by the Minerva Foundation.\\

S.M.\ and A.D.\ contributed equally to this work.

\end{acknowledgements}

\bibliography{bibfile}	

\begin{thebibliography}{80}
\bibitem [{ave()}]{averagedef}%
  \BibitemOpen
  \href@noop {} {}\bibinfo {note} {We define the time average of an operator
  $x$ as $\langle x \rangle = \lim_{\tau \rightarrow \infty} \frac{1}{\tau}
  \int_{-\tau}^{\tau} x(t) dt $}\BibitemShut {NoStop}%
\end{thebibliography}
\end{document}


\title{\textbf{Supplemental material for: ``Full Classification of Transport on an Equilibrated 5/2 Edge via Shot Noise"}}

\author{Sourav Manna}
\affiliation{Department of Condensed Matter Physics, Weizmann Institute of Science, Rehovot 7610001, Israel}
\affiliation{Raymond and Beverly Sackler School of Physics and Astronomy, Tel-Aviv University, Tel Aviv 6997801, Israel}
	
\author{Ankur Das}
\affiliation{Department of Condensed Matter Physics, Weizmann Institute of Science, Rehovot 7610001, Israel}
	
\author{Moshe Goldstein}
\affiliation{Raymond and Beverly Sackler School of Physics and Astronomy, Tel-Aviv University, Tel Aviv 6997801, Israel}
	
\author{Yuval Gefen}
\affiliation{Department of Condensed Matter Physics, Weizmann Institute of Science, Rehovot 7610001, Israel}

\begin{abstract}
The supplemental material contains the following details:
\begin{enumerate}

    \item Proposed device and configurations and general expression for the current-current correlations (CCC) in \cref{device_and_Noise}. 

    \item Expressions for the CCC for three specific cases of edge equilibrations in the device in \cref{CCC_eq}.

    \item CCC values for specific cases with Pfaffian (Pf), anti-Pf (APf), and particle-hole-Pf (PHPf) states in different configurations in \cref{CCCvalues}. 
\end{enumerate}
\end{abstract}

\maketitle

\section{Device and general expression of CCC}\label{device_and_Noise}
In this section we re-describe our proposed device in \cref{device} to facilitate the follow up discussions for the readers. In \cref{Noise} we derive general expression for the CCC in our device.

\subsection{Proposed device}\label{device}
\begin{figure}[h!]
	\includegraphics[width=\columnwidth]{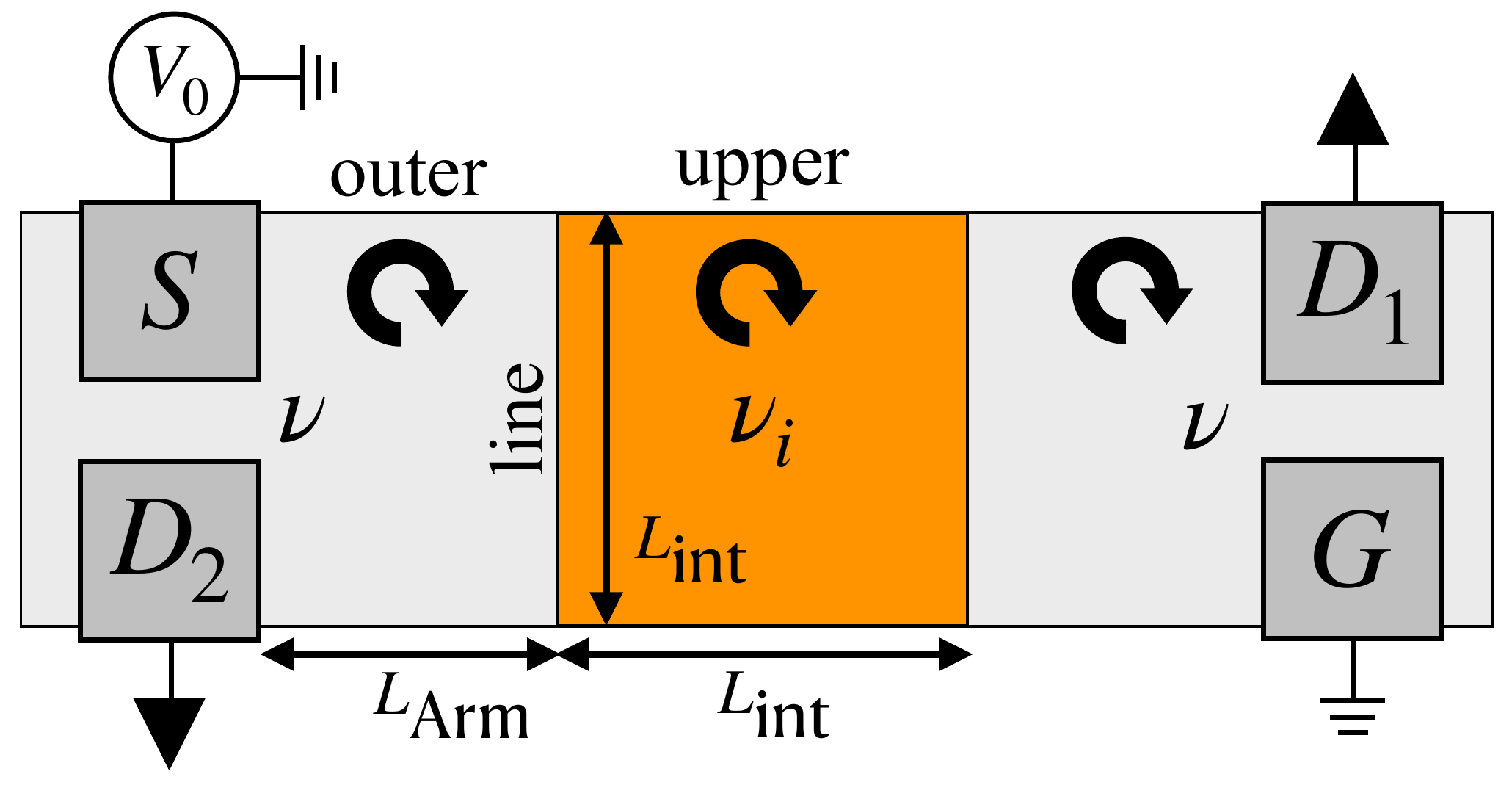}
	\caption{Our proposed device (same as in main text): A Hall bar with filling $\nu$ with the middle region depleted to $\nu_i(<\nu)$.
	This creates two interfaces (``line"). We choose $\{\nu, \nu_i\}$ as $\{5/2,2\}$ or $\{5/2,7/3\}$ or $\{3,5/2\}$. We also have the boundary of the vacuum with $\nu$ (``outer") and with $\nu_i$ (``upper"). There are four contacts: a source $S$ (biased by a dc voltage $V_0$), a ground $G$, and two drains
	$D_1, D_2$. Each circular arrow indicates the chirality of charge
    propagation of the respective filling. $L_{\text{Arm}}$ and $L_{\text{int}}$, the geometric lengths, are assumed to be of the same order of magnitude.}\label{Dev}
\end{figure}

Our device is a Hall bar with filling $\nu$ and a depleted middle region with filling $\nu_i(<\nu)$, c.f.\ \cref{Dev}. We choose
$\{\nu, \nu_i\}$ as $\{5/2,2\}$ or $\{5/2,7/3\}$ or $\{3,5/2\}$.
We have a source contact $S$ which is biased by a dc voltage $V_0$, a ground contact $G$, and two drains $D_1, D_2$.

\subsection{General expression for the CCC}\label{Noise}
\begin{figure}[h!]
	\includegraphics[width=\columnwidth]{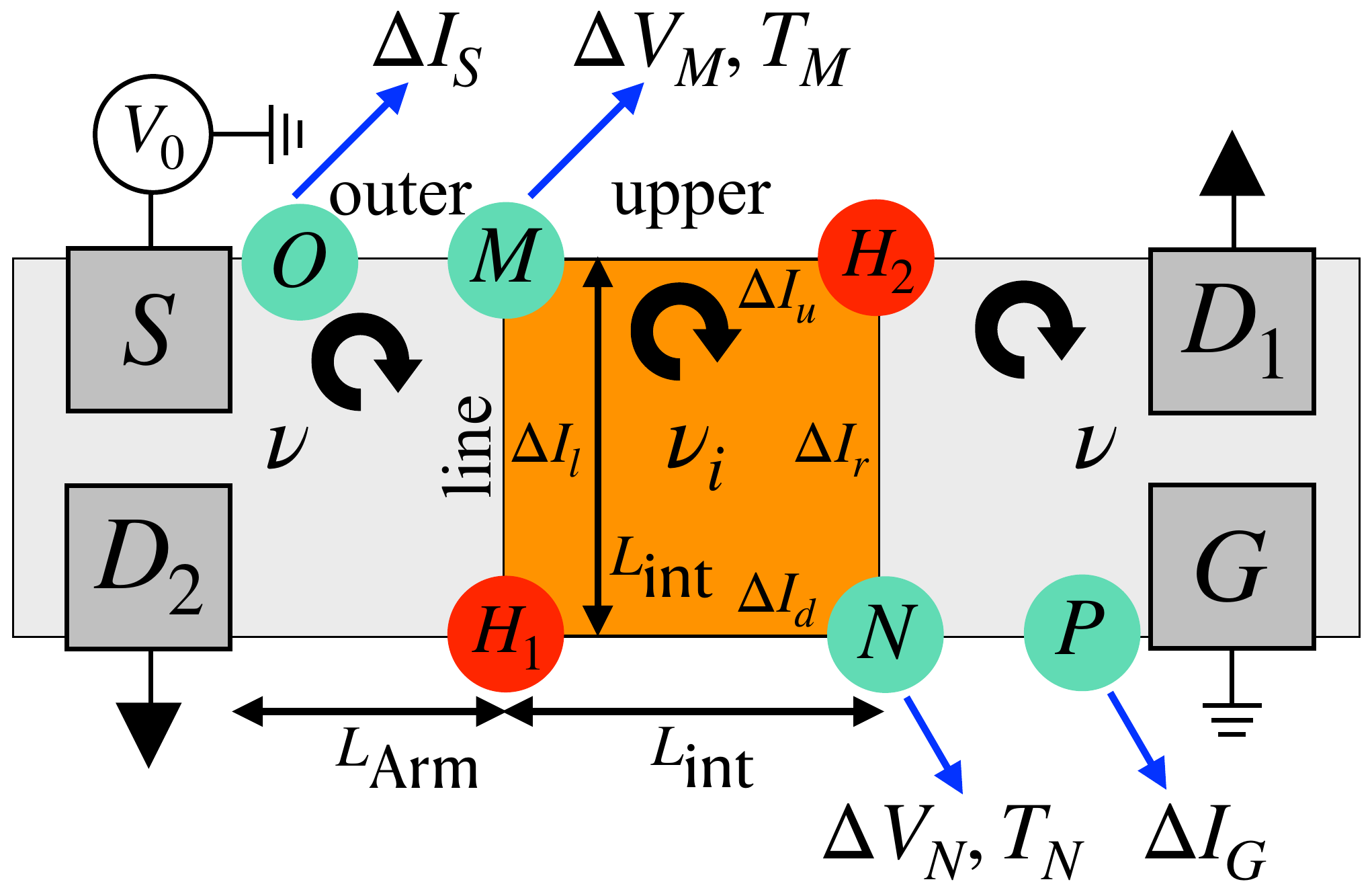}
	\caption{Noise generation in our proposed device (see \cref{Dev}), where charge and heat are equilibrated in each segment. Each circular arrow shows the chirality of charge propagation of the respective
	filling.  Voltage drops occur at the hot spots $H_1,H_2$ (red circles), resulting in noise spots $M,N,O,P$ (green circles) \cite{PhysRevB.101.075308}. 
	The current fluctuations $\Delta I_S$ and $\Delta I_G$ are at $O$ and $P$, respectively. We denote fluctuations in the local voltages at the noise spots $M$ and $N$
	by $\Delta V_M$ and $\Delta V_N$, respectively. The temperatures at the noise spots $M$ and $N$ are $T_M$ and $T_N$, respectively. We denote the current fluctuations in the segments of the middle region, with filling $\nu_i$, as $\Delta I_u, \Delta I_d, \Delta I_l, \Delta I_r$. } \label{IncoherentSM}
\end{figure}
Let us denote by $I_1$ and $I_2$ the currents entering the drains $D_1$ and $D_2$, respectively (\cref{IncoherentSM}). The corresponding current-current auto-correlations are defined as $\delta^2 I_1=\langle (I_1 - \langle I_1 \rangle)^2
\rangle$ in $D_1$ and $\delta^2 I_2=\langle (I_2 - \langle I_2 \rangle)^2 \rangle$ in $D_2$, while the cross-correlation is $\delta^2 I_c=\left\langle \left(I_1 - \langle I_1 \rangle\right)
\left(I_2 - \langle I_2 \rangle\right) \right\rangle$, where we define the time average of an operator $x$ as $\langle x \rangle = \lim_{\tau \rightarrow \infty} \frac{1}{\tau} \int_{-\tau}^{\tau} x(t) dt $. We consult \cref{IncoherentSM} and follow Ref.\ \cite{PhysRevB.101.075308} to write 
\begin{equation}\label{I1I2fluc}
	\begin{split}
		(I_1 - \langle I_1 \rangle) &= \Delta I_u + \Delta I_r,\\ (I_2 - \langle I_2 \rangle) &= \Delta I_d + \Delta I_l.
	\end{split}
\end{equation}
The current fluctuations $\Delta I_u, \Delta I_d, \Delta I_l, \Delta I_r$, are composed of the voltage fluctuations,
$\Delta V_M, \Delta V_N$, and the thermal fluctuations $\Delta I_u^{\text{th}}, \Delta I_d^{\text{th}}, \Delta I_l^{\text{th}}, \Delta I_r^{\text{th}}$. We write
\begin{equation}
	\begin{split}
		& \Delta I_l = (\nu-\nu_i) \frac{e^2}{h} \Delta V_M + \Delta I_l^{\text{th}},\\& \Delta I_u = \nu_i \frac{e^2}{h} \Delta V_M + \Delta I_u^{\text{th}},\\&\Delta I_r = (\nu-\nu_i) \frac{e^2}{h} \Delta V_N + \Delta I_r^{\text{th}},\\& \Delta I_d = \nu_i \frac{e^2}{h} \Delta V_N + \Delta I_r^{\text{th}},
	\end{split}
\end{equation}
where $e$ is the electron charge and $h$ is the Planck's constant. The voltage fluctuations are
\begin{equation}
	\begin{split}
		&\nu\frac{e^2}{h}\Delta V_M = \Delta I_S - \Delta I_l^{\text{th}} - \Delta I_u^{\text{th}},\\&\nu\frac{e^2}{h}\Delta V_N = \Delta I_G - \Delta I_r^{\text{th}} - \Delta I_d^{\text{th}}.
	\end{split}
\end{equation}
From \cref{I1I2fluc} we find
\begin{equation}
	\begin{split}
		(I_1 - \langle I_1 \rangle) &= \nu_i \frac{e^2}{h} \Delta V_M + \Delta I_u^{\text{th}}+(\nu-\nu_i) \frac{e^2}{h} \Delta V_N + \Delta I_r^{\text{th}}\\&=\frac{\nu_i}{\nu}\Delta I_S +
		\frac{(\nu-\nu_i)}{\nu}\Delta I_G \\&+ \frac{(\nu-\nu_i)}{\nu}(\Delta I_u^{\text{th}} - \Delta I_d^{\text{th}})+\frac{\nu_i}{\nu}(\Delta I_r^{\text{th}} - \Delta I_l^{\text{th}}),
	\end{split}
\end{equation}
and
\begin{equation}
	\begin{split}
		(I_2 - \langle I_2 \rangle) &=\frac{\nu_i}{\nu}\Delta I_G + \frac{(\nu-\nu_i)}{\nu}\Delta I_S \\&+
		\frac{(\nu-\nu_i)}{\nu}(\Delta I_d^{\text{th}} - \Delta I_u^{\text{th}})+\frac{\nu_i}{\nu}(\Delta I_l^{\text{th}} - \Delta I_r^{\text{th}}).
	\end{split}
\end{equation}
We obtain the expressions of $\delta^2 I_1, \delta^2 I_2$ and $\delta^2 I_c$ as
\begin{equation}\label{auto1}
	\begin{split}
		\delta^2 I_1 &=2\Bigg(\frac{e^2}{h}\Bigg)\frac{\nu_i}{\nu}(\nu-\nu_i)k_{\text{B}} (T_M+T_N) \\&+ \frac{1}{\nu^2}\Big[\nu_i^2 \langle (\Delta I_S)^2 \rangle + (\nu-\nu_i )^2\langle (\Delta I_G)^2 \rangle\Big],
	\end{split}
\end{equation}
\begin{equation}\label{auto2}
	\begin{split}
		\delta^2 I_2 &=2\Bigg(\frac{e^2}{h}\Bigg)\frac{\nu_i}{\nu}(\nu-\nu_i)k_{\text{B}} (T_M+T_N) \\&+
		\frac{1}{\nu^2}\Big[\nu_i^2 \langle (\Delta I_G)^2 \rangle + (\nu-\nu_i )^2\langle (\Delta I_S)^2 \rangle\Big],
	\end{split}
\end{equation}
and
\begin{equation}\label{cross}
	\begin{split}
		\delta^2 I_c &=-2\Bigg(\frac{e^2}{h}\Bigg)\frac{\nu_i}{\nu}(\nu-\nu_i)k_{\text{B}} (T_M+T_N) \\&+ \frac{\nu_i(\nu-\nu_i)}{\nu^2}\Big[\langle (\Delta I_G)^2 \rangle + \langle (\Delta I_S)^2 \rangle\Big],
	\end{split}
\end{equation}
where $T_M, T_N$ are, respectively, the temperatures at the noise spots $M$ and $N$. 
To derive \cref{auto1,auto2,cross} we have used the local Johnson-Nyquist relations for thermal noise,
\begin{equation}
	\begin{split}
		&\ \ \ \ \  \langle (\Delta I_l^{\text{th}})^2 \rangle = \frac{2 e^2}{h} (\nu-\nu_i) k_{\text{B}} T_M,\\& \ \ \ \ \ \langle (\Delta I_u^{\text{th}})^2 \rangle = \frac{2 e^2}{h} \nu_i k_{\text{B}} T_M,\\& \ \ \ \ \ \langle (\Delta I_r^{\text{th}})^2 \rangle = \frac{2 e^2}{h}
		(\nu-\nu_i)k_{\text{B}} T_N,\\& \ \ \ \ \ \langle (\Delta I_d^{\text{th}})^2 \rangle = \frac{2 e^2}{h} \nu_i k_{\text{B}} T_N,\\& \langle (\Delta I_i^{\text{th}} \Delta I_j^{\text{th}}) \rangle = 0,\ \text{for}\ i \neq j\ \text{and}\ i,j \in \{l,u,r,d\},
	\end{split}
\end{equation}
where $k_{\text{B}}$ is the Boltzmann constant. The Fano factors are then defined as
\begin{equation}
	\begin{split}
		F_j=\frac{|\delta^2 I_j|}{2e \langle I\rangle t (1-t)},
	\end{split}
\end{equation}
where $j \in \{1,2,c\}$, $I$ is the source current and $t=\langle I_1 \rangle/\langle I \rangle$.

\section{Expression for the CCC for three specific cases of edge equilibrations}\label{CCC_eq}
We assume that the charge is fully equilibrated, hence charge transport is ballistic, moving ``downstream" along each segment of the device (\cref{Dev}). We call the direction opposite to charge flow (``upstream") antiballistic. For the cases
we consider, heat is also equilibrated and heat transport 
can be either ballistic or antiballistic in each segment of the device. Specifically we calculate CCC for the following three cases:

$1)$ heat transport is ballistic in ``outer", antiballistic in ``line", and ballistic in
``upper" segments, 

$2)$ heat transport is antiballistic in ``outer", antiballistic in ``line", and ballistic in
``upper" segments,

$3)$ heat transport is ballistic in ``outer", ballistic in ``line", and antiballistic in
``upper" segments.

\subsection{CCC when heat transport is ballistic in ``outer", antiballistic in ``line", and ballistic in
``upper" segments}\label{BABB}
\begin{figure}[h!]
	\includegraphics[width=\columnwidth]{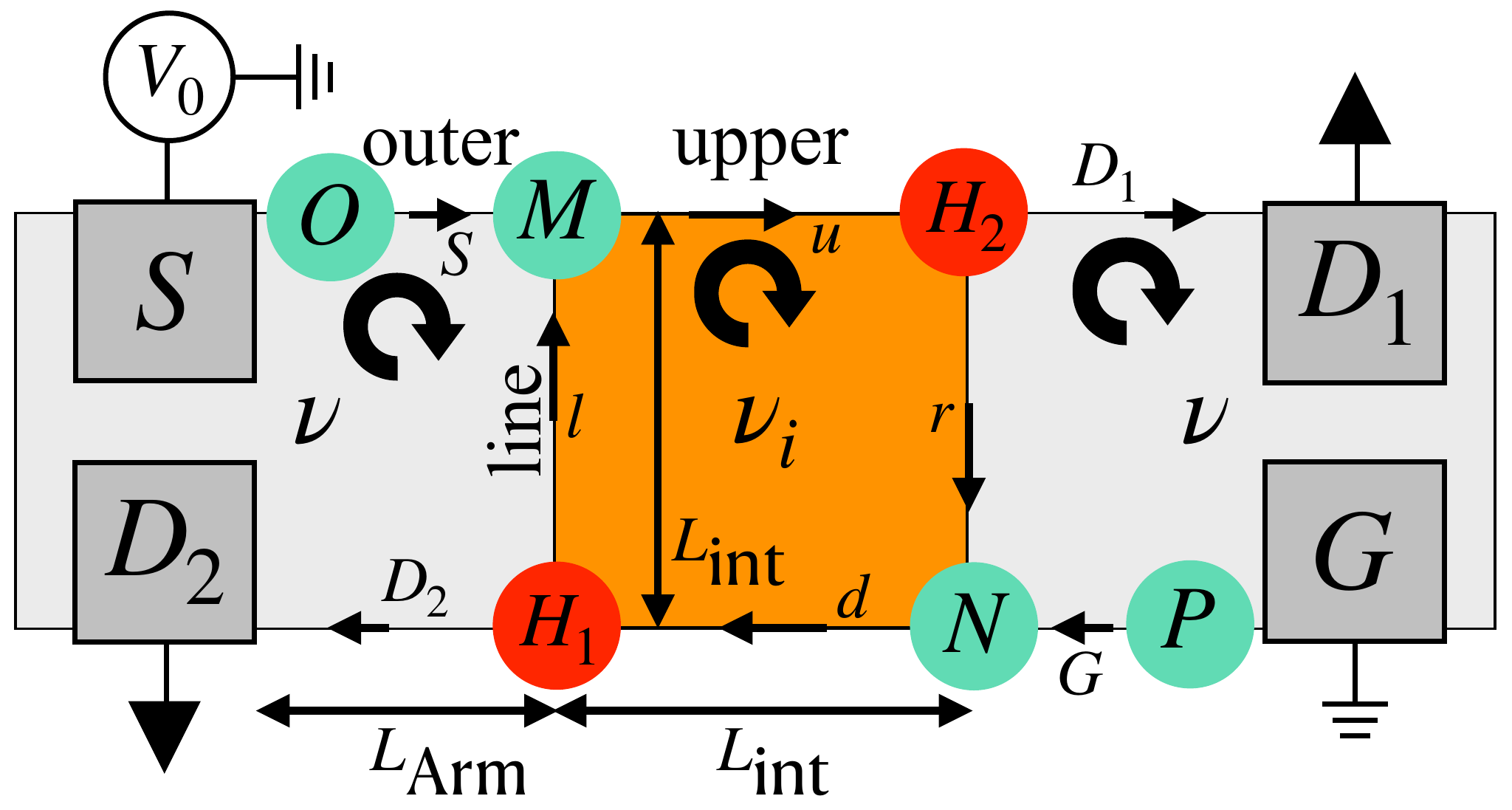}
	\caption{Noise generation in our proposed device (see \cref{Dev}): Charge is fully equilibrated, hence charge transport is ballistic, moving ``downstream" along each segment of the device. Each circular arrow shows the chirality of charge
propagation of the respective filling. We call the direction opposite to charge flow (``upstream") antiballistic. Heat is also
equilibrated and the heat transport is ballistic in ``outer", antiballistic in ``line", and ballistic in
``upper" segments, as shown by the arrows. Voltage drops occur at the hot spots $H_1,H_2$ (red circles), resulting in the noise spots $M,N,O,P$ (green circles) \cite{PhysRevB.101.075308}.
} \label{heat}
\end{figure}
We consult \cref{heat} and derive expressions of CCC 
when heat transport is ballistic in ``outer", antiballistic in ``line", and ballistic in
``upper" segments.

\subsubsection{Temperature calculation}
We refer to \cref{heat}. Here we calculate the expression of the temperatures $T_M, T_N$ at
the noise spots $M, N$, respectively. We assume zero temperature in the contacts/leads. Conservation of energy at $H_1, N, H_2, M$ implies the following expressions \cite{PhysRevB.101.075308}:
\begin{equation}\label{heatcurrent}
	\begin{split}
		&J_{D_2} + J_l = P_{H_1} + J_d,\\&J_G+J_r=J_d,\\&J_{D_1}+J_r=P_{H_2}+J_u,\\&J_S+J_l=J_u,
	\end{split}
\end{equation}
where $J_i$ is the heat current along $i \in \{ S, G, D_1, D_2, u,r,d,l \}$th segment, and $P_{H_1}, P_{H_2}$ are, respectively, the dissipated powers in the hot spots $H_1, H_2$. We write
\begin{equation}\label{heatcurrentBABB}
	\begin{split}
		&J_S=0,\ J_u=\frac{\kappa}{2}T_M^2(\delta c)_u,\ J_{D_1}=\frac{\kappa}{2}T_{H_2}^2(\delta c)_1,\\&J_G=0,\ J_d=\frac{\kappa}{2}T_N^2(\delta c)_d,\ J_{D_2}=\frac{\kappa}{2}T_{H_1}^2(\delta c)_2,\\&J_l=\frac{\kappa}{2}T_{H_1}^2(\delta c)_l,\ J_r=\frac{\kappa}{2}T_{H_2}^2(\delta c)_r,
	\end{split}
\end{equation}
where 
\begin{equation}
	\begin{split}
		\kappa = \frac{\pi^2 k_{\text{B}}^2}{3h},\ P_{H_1}=P_{H_2}=\frac{e^2 V_0^2}{h} \frac{(\nu-\nu_i)\nu_i}{2\nu},
	\end{split}
\end{equation}
and $T_M, T_N, T_{H_1}, T_{H_2}$ are the temperatures at
$M, N, H_1, H_2$ respectively.
The dissipated powers $P_{H_1}, P_{H_2}$ arise due to the Joule heating of the associated voltage drop at $H_1, H_2$, respectively. We obtain $P_{H_1}=P_{H_2}$, since the voltage drop only depends on $\nu$ and $\nu_i$ \cite{PhysRevB.101.075308}. Here $(\delta c)_i = |(c_{\text{down}}-c_{\text{up}})_i|$ is the modulus of difference between the central charges of downstream and upstream modes in the $i$th segment 
of the device. 

Taking all these into account we find
\begin{equation}
	\begin{split}
		&\frac{\kappa}{2}T_{H_1}^2(\delta c)_2+\frac{\kappa}{2}T_{H_1}^2(\delta c)_l=P_{H_1}+\frac{\kappa}{2}T_N^2(\delta c)_d,\\&0+\frac{\kappa}{2}T_{H_2}^2(\delta c)_r=\frac{\kappa}{2}T_N^2(\delta c)_d,\\&\frac{\kappa}{2}T_{H_2}^2(\delta c)_1+\frac{\kappa}{2}T_{H_2}^2(\delta c)_r=P_{H_2}+\frac{\kappa}{2}T_M^2(\delta c)_u,\\&0+\frac{\kappa}{2}T_{H_1}^2(\delta c)_l=\frac{\kappa}{2}T_M^2(\delta c)_u,
	\end{split}
\end{equation}
and hence,
\begin{equation}
	\begin{split}
		&\frac{[(\delta c)_1 + (\delta c)_r]}{(\delta c)_r}(\delta c)_d\frac{\kappa}{2}T_N^2=P_{H_2}+\frac{\kappa}{2}(\delta c)_uT_M^2,\\&\frac{[(\delta c)_2 + (\delta c)_l]}{(\delta c)_l}(\delta c)_u\frac{\kappa}{2}T_M^2=P_{H_1}+\frac{\kappa}{2}(\delta c)_dT_N^2.
	\end{split}
\end{equation}
Therefore,
\begin{equation}
	\begin{split}
		&\frac{\kappa}{2}(\delta c)_uT_M^2\Bigg[ 1+\frac{(\delta c)_2+(\delta c)_l}{(\delta c)_l} \Bigg] \\&= \frac{\kappa}{2}(\delta c)_dT_N^2\Bigg[ 1+\frac{(\delta c)_1+(\delta c)_r}{(\delta c)_r} \Bigg].
	\end{split}
\end{equation}
We note that $(\delta c)_u=(\delta c)_d,\ (\delta c)_2=(\delta c)_1,\ (\delta c)_l=(\delta c)_r$, hence $T_M=T_N$. Thus we find
\begin{equation}
	\begin{split}
		\frac{\kappa T_M^2}{2}\frac{(\delta c)_1}{(\delta c)_r}(\delta c)_d=P_{H_2}=\frac{e^2 V_0^2}{h} \frac{(\nu-\nu_i)\nu_i}{2\nu},
	\end{split}
\end{equation}
and, finally,
\begin{equation}\label{TcBABB}
	\begin{split}
		T_M=T_N=\frac{eV_0}{\pi k_{\text{B}}}\sqrt{\frac{3(\nu-\nu_i)\nu_i}{\nu} \frac{(\delta c)_r}{(\delta c)_1 (\delta c)_d}}.
	\end{split}
\end{equation}

\subsubsection{Noise spot contributions}\label{BABBIs}
We refer to \cref{heat}. Here we calculate the contributions $\langle (\Delta I_S)^2 \rangle,\ \langle (\Delta I_G)^2 \rangle$
of the noise spots $O, P$, respectively. To evaluate $\langle (\Delta I_S)^2 \rangle,\ \langle (\Delta I_G)^2 \rangle$ we need to compute the following integral \cite{PhysRevB.101.075308},
\begin{equation}\label{outernoise}
	\begin{split}
		&\langle (\Delta I_S)^2 \rangle = \langle (\Delta I_G)^2 \rangle \\&= \frac{2e^2}{h l_{\text{eq}}^{\text{ch}}}\nu \frac{\nu_{-}}{\nu_{+}}\bigintss_0^{L_{\text{Arm}}}dx \frac{e^{\frac{-2x}{l_{\text{eq}}^{\text{ch}}}}k_{\text{B}}[T_{+}(x)+T_{-}(x)]}{[1-(e^{\frac{-L_{\text{Arm}}}{l_{\text{eq}}^{\text{ch}}}}\frac{\nu_{-}}{\nu_{+}})]^2},
	\end{split}
\end{equation}
where $l_{\text{eq}}^{\text{ch}}$ is the charge equilibration length, $\nu=(\nu_{+}-\nu_{-})$, and  $T_{\pm}(x)$ are the temperature profiles in $\nu_{\pm}$, respectively. We note that $T_{\pm}(x)$ depend on the thermal equilibration length $l_{\text{eq}}$, to be further discussed below. Here $\nu_{+}$ is the filling factor of the downstream
mode and $\nu_{-}$ is the filling factor of the upstream mode.

To compute $T_{\pm}(x)$ we solve the following differential equation for the local temperatures,
\begin{equation}
	\begin{split}
		\begin{bmatrix}
			\partial_x T^2_{+}\\
			\partial_x T^2_{-}
		\end{bmatrix}=\frac{\gamma}{l}
		\begin{bmatrix}
			-c_{\text{up}} & c_{\text{up}}\\
			-c_{\text{down}} & c_{\text{down}}
		\end{bmatrix}
		\begin{bmatrix}
			T^2_{+}\\
			T^2_{-}
		\end{bmatrix},
	\end{split}
\end{equation}
where $l = \frac{(\nu_{+}-\nu_{-})l_{\text{eq}}}{\nu_{+}\nu_{-}}$ and $c_{\text{down}}\ (c_{\text{up}})$ is the central charge of the downstream (upstream) mode. Here $\gamma$ is a parameter of order unity, characterizing the deviation of the ratio of intermode charge and heat tunneling conductances
from Wiedemann-Franz law \cite{PhysRevB.101.075308}, where the Wiedemann-Franz law corresponds to $\gamma=1$. We assume that there are no voltage drops along the ``outer" segment (\cref{heat}), hence there is no Joule heating contribution in the following \cite{PhysRevB.101.075308}.
The boundary conditions are
\begin{equation}
	\begin{split}
		T_{+}(0)=0,\ T_{-}(L_{\text{Arm}})=T_M.
	\end{split}
\end{equation}
The temperature profiles are found to be 
\begin{equation}\label{temp}
	\begin{split}
		&T^2_{+}=T_M^2\Bigg( \frac{c_{\text{up}}-c_{\text{up}}e^{-\alpha x/l_{\text{eq}}}}{c_{\text{up}}-c_{\text{down}}e^{-\alpha L_{\text{Arm}}/l_{\text{eq}}}} \Bigg),\\&T^2_{-}=T_M^2\Bigg( \frac{c_{\text{up}}-c_{\text{down}}e^{-\alpha x/l_{\text{eq}}}}{c_{\text{up}}-c_{\text{down}}e^{-\alpha L_{\text{Arm}}/l_{\text{eq}}}} \Bigg),
	\end{split}
\end{equation}
where $\alpha = \frac{-(c_{\text{down}}-c_{\text{up}})\gamma \nu_{+}\nu_{-}}{\nu}$. Plugging the temperature profiles into
\cref{outernoise}, we can
derive the expression of $\langle (\Delta I_S)^2 \rangle = \langle (\Delta I_G)^2 \rangle$.
We note that for full thermal equilibration with length $l_{\text{eq}}^{\text{full}}$, we have $l_{\text{eq}}=l_{\text{eq}}^{\text{full}}$, while for partial thermal equilibration with length $l_{\text{eq}}^{\text{par}}$, we have $l_{\text{eq}}=l_{\text{eq}}^{\text{par}}$. We are interested in the limit where $L_{\text{Arm}} \gg l_{\text{eq}}$ and $L_{\text{Arm}} \gg l_{\text{eq}}^{\text{ch}}$.

\subsection{CCC when heat transport is antiballistic in ``outer", antiballistic in ``line", and ballistic in
``upper" segments}\label{ABABB}
\begin{figure}[h!]
	\includegraphics[width=\columnwidth]{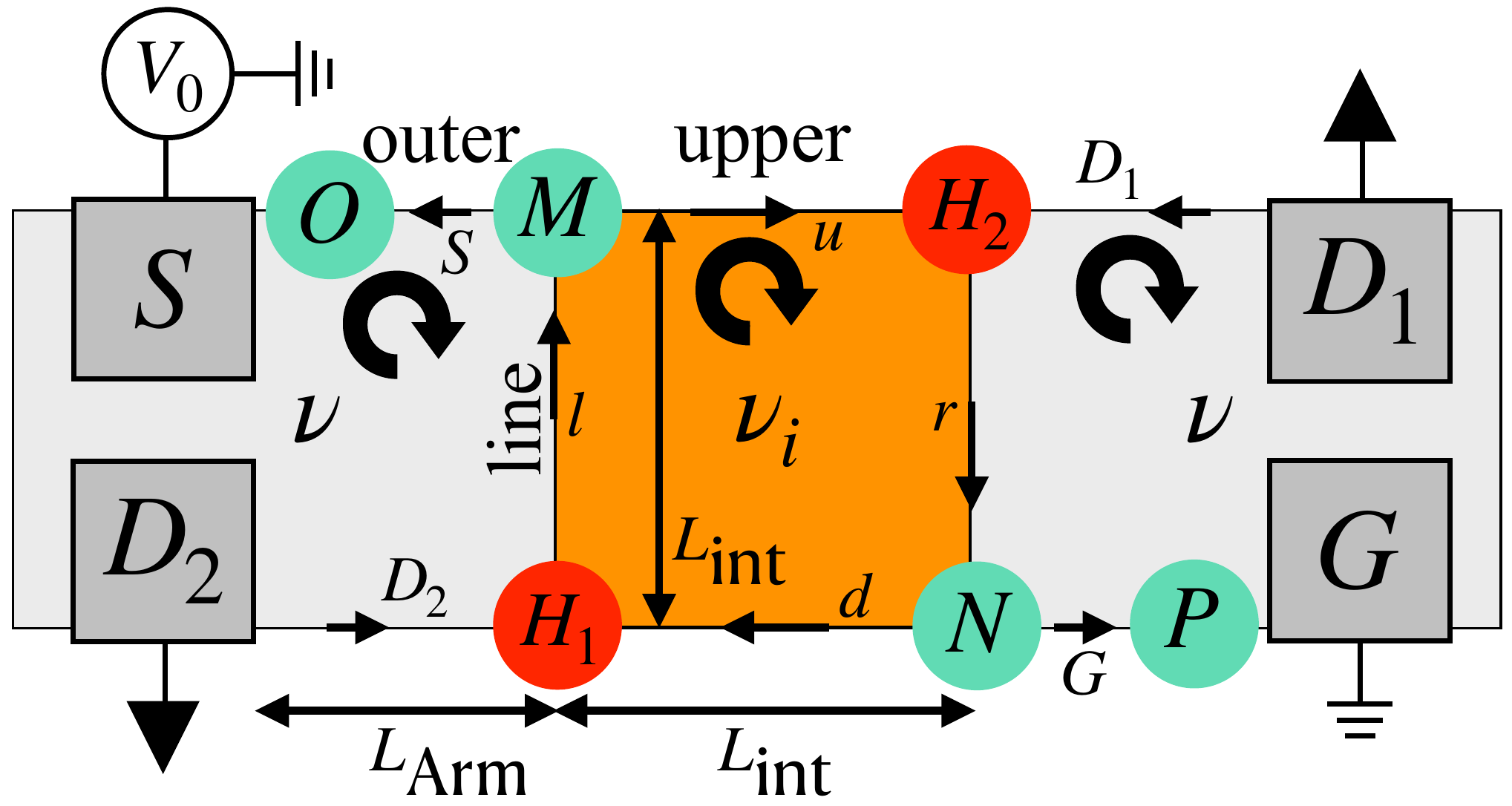}
	\caption{Noise generation in our proposed device (see \cref{Dev}): Charge is fully equilibrated, hence charge transport is ballistic, moving ``downstream" along each segment of the device. Each circular arrow shows the chirality of charge
propagation of the respective filling. We call the direction opposite to charge flow (``upstream") antiballistic. Heat is also
equilibrated and the heat transport is antiballistic in ``outer", antiballistic in ``line", and ballistic in
``upper" segments, as shown by the arrows. Voltage drops occur at the hot spots $H_1,H_2$ (red circles), resulting in noise spots $M,N,O,P$ (green circles) \cite{PhysRevB.101.075308}.
} \label{heatABABB}
\end{figure}
We consult \cref{heatABABB} and derive expressions for the CCC 
when heat transport is antiballistic in ``outer", antiballistic in ``line", and ballistic in
``upper" segments.

\subsubsection{Temperature calculation}
We refer to \cref{heatABABB}. Here we calculate the expression for the temperatures $T_M, T_N$ at
the noise spots $M, N$, respectively. We assume zero temperature in the contacts/leads. Conservation of energy at $H_1, N, H_2, M$ implies the following expressions
\cite{PhysRevB.101.075308}:
\begin{equation}\
	\begin{split}
		& J_l = J_{D_2} +P_{H_1} + J_d,\\&J_G+J_d=J_r,\\&J_r=J_{D_1}+P_{H_2}+J_u,\\&J_S+J_u=J_l,
	\end{split}
\end{equation}
where $J_i$ is the heat current along $i \in \{ S, G, D_1, D_2, u,r,d,l \}$th segment, and $P_{H_1}, P_{H_2}$ are, respectively, the dissipated powers in the hot spots $H_1, H_2$. We write
\begin{equation}
	\begin{split}
		&J_S=\frac{\kappa}{2}T_M^2(\delta c)_S,\ J_u=\frac{\kappa}{2}T_M^2(\delta c)_u,\ J_{D_1}=0,\\&J_G=\frac{\kappa}{2}T_N^2(\delta c)_G,\ J_d=\frac{\kappa}{2}T_N^2(\delta c)_d,\ J_{D_2}=0,\\&J_l=\frac{\kappa}{2}T_{H_1}^2(\delta c)_l,\ J_r=\frac{\kappa}{2}T_{H_2}^2(\delta c)_r,
	\end{split}
\end{equation}
where 
\begin{equation}
	\begin{split}
		\kappa = \frac{\pi^2 k_{\text{B}}^2}{3h},\ P_{H_1}=P_{H_2}=\frac{e^2 V_0^2}{h} \frac{(\nu-\nu_i)\nu_i}{2\nu},
	\end{split}
\end{equation}
and $T_M, T_N, T_{H_1}, T_{H_2}$ are the temperatures at
$M, N, H_1, H_2$, respectively.
The dissipated powers $P_{H_1}, P_{H_2}$ arise due to the Joule heating of the associated voltage drop at $H_1, H_2$, respectively. We obtain $P_{H_1}=P_{H_2}$, since the voltage drop only depends on $\nu$ and $\nu_i$ \cite{PhysRevB.101.075308}. Here $(\delta c)_i = |(c_{\text{down}}-c_{\text{up}})_i|$ is the modulus of difference between the central charges of downstream and upstream modes in the $i$th segment 
of the deivce. 

With all these in mind we find 
\begin{equation}
	\begin{split}
		&\frac{\kappa}{2}T_{H_1}^2(\delta c)_l=0+P_{H_1}+\frac{\kappa}{2}T_N^2(\delta c)_d,\\&\frac{\kappa}{2}T_N^2(\delta c)_G+\frac{\kappa}{2}T_N^2(\delta c)_d=\frac{\kappa}{2}T_{H_2}^2(\delta c)_r,\\&\frac{\kappa}{2}T_{H_2}^2(\delta c)_r=0+P_{H_2}+\frac{\kappa}{2}T_M^2(\delta c)_u,\\&\frac{\kappa}{2}T_M^2(\delta c)_S+\frac{\kappa}{2}T_M^2(\delta c)_u=\frac{\kappa}{2}T_{H_1}^2(\delta c)_l,
	\end{split}
\end{equation}
and thus
\begin{equation}
	\begin{split}
		&[(\delta c)_G + (\delta c)_d]\frac{\kappa}{2}T_N^2=P_{H_2}+\frac{\kappa}{2}(\delta c)_uT_M^2,\\&[(\delta c)_S + (\delta c)_u]\frac{\kappa}{2}T_M^2=P_{H_1}+\frac{\kappa}{2}(\delta c)_dT_N^2.
	\end{split}
\end{equation}
Therefore,
\begin{equation}
	\begin{split}
		T_M^2=\frac{[(\delta c)_G + 2(\delta c)_d]}{[(\delta c)_S + 2(\delta c)_u]}T_N^2.
	\end{split}
\end{equation}
We note that $(\delta c)_u=(\delta c)_d,\ (\delta c)_2=(\delta c)_1,\ (\delta c)_l=(\delta c)_r,\ (\delta c)_S=(\delta c)_G$, hence $T_M=T_N$. Thus we find
\begin{equation}\label{TcABABB}
	\begin{split}
		T_M=T_N=\frac{eV_0}{\pi k_{\text{B}}}\sqrt{\frac{3(\nu-\nu_i)\nu_i}{\nu} \frac{1}{(\delta c)_G}}.
	\end{split}
\end{equation}

\subsubsection{Noise spot contributions}
To calculate the contributions $\langle (\Delta I_S)^2 \rangle,\ \langle (\Delta I_G)^2 \rangle$
from the noise spots $O, P$, respectively,
we can use the same equations as in \cref{BABBIs}.

\subsection{CCC when heat transport is ballistic in ``outer", ballistic in ``line", and antiballistic in
``upper" segments}\label{BBAB}
\begin{figure}[h!]
	\includegraphics[width=\columnwidth]{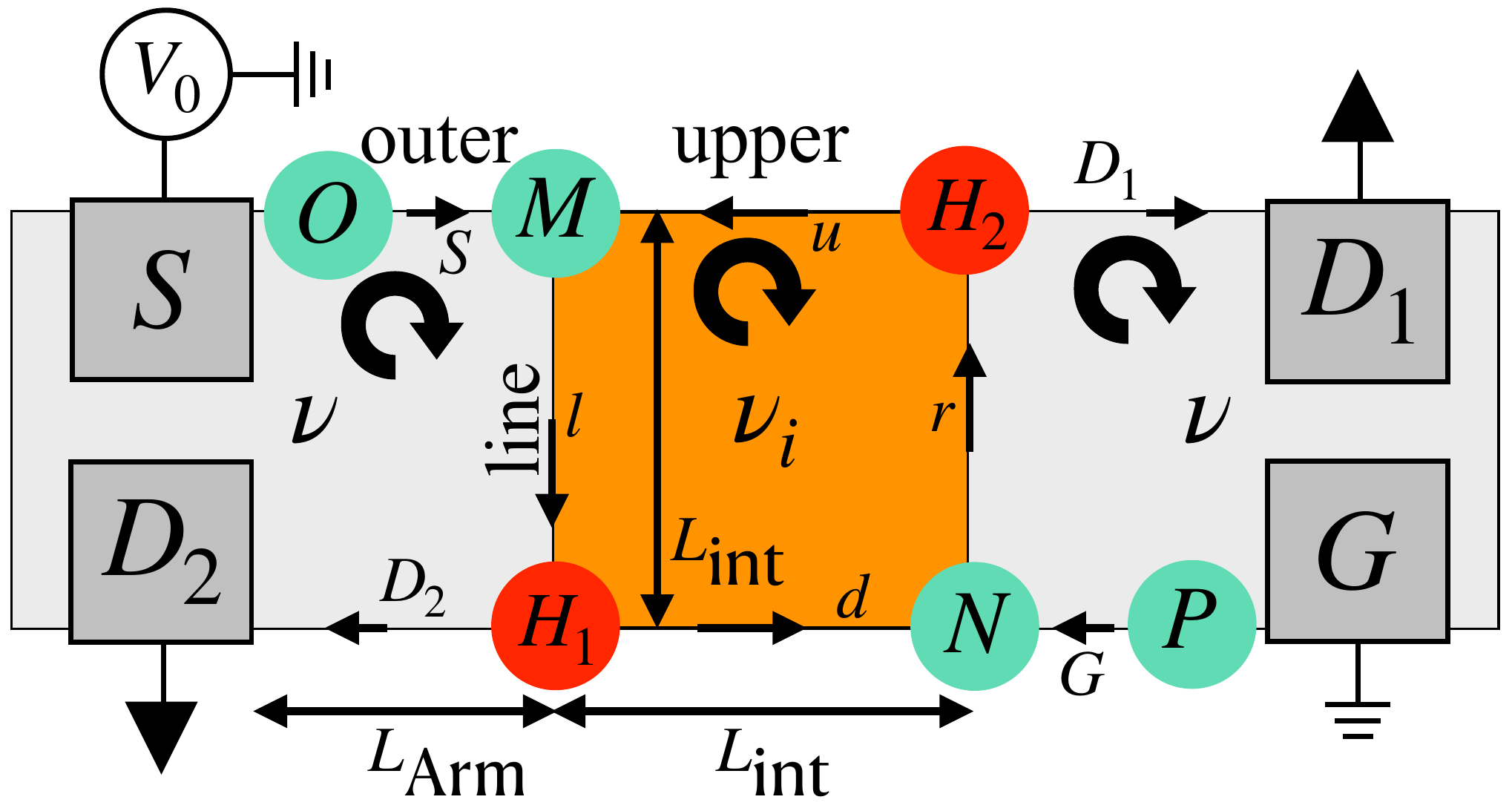}
	\caption{Noise generation in our proposed device (see \cref{Dev}): Charge is fully equilibrated, hence charge transport is ballistic, moving ``downstream" along each segment of the device. Each circular arrow shows the chirality of charge
propagation of the respective filling. We call the direction opposite to charge flow (``upstream") antiballistic. Heat is also
equilibrated and the heat transport is ballistic in ``outer", ballistic in ``line", and antiballistic in
``upper" segments, as shown by the arrows. Voltage drops occur at the hot spots $H_1,H_2$ (red circles), resulting in noise spots $M,N,O,P$ (green circles) \cite{PhysRevB.101.075308}.
} \label{heatBBAB}
\end{figure}
We consult \cref{heatBBAB} and derive expressions for the CCC 
when heat transport is ballistic in ``outer", ballistic in ``line", and antiballistic in
``upper" segments.

\subsubsection{Temperature calculation}
We refer to \cref{heatBBAB}. Here we calculate the expression of the temperatures $T_M, T_N$ at
the noise spots $M, N$, respectively. We assume zero temperature in the contacts/leads. Conservation of energy at $H_1, N, H_2, M$ implies the following expressions
\cite{PhysRevB.101.075308}:
\begin{equation}
	\begin{split}
		&J_{D_2} + J_d = P_{H_1} + J_l,\\&J_G+J_d=J_r,\\&J_{D_1}+J_u=P_{H_2}+J_r,\\&J_S+J_u=J_l,
	\end{split}
\end{equation}
where $J_i$ is the heat current along $i \in \{ S, G, D_1, D_2, u,r,d,l \}$th segment, and $P_{H_1}, P_{H_2}$ are, respectively, the dissipated powers in the hot spots $H_1, H_2$. We write
\begin{equation}\label{heatcurrentBABB2}
	\begin{split}
		&J_S=0,\ J_u=\frac{\kappa}{2}T_{H_2}^2(\delta c)_u,\ J_{D_1}=\frac{\kappa}{2}T_{H_2}^2(\delta c)_1,\\&J_G=0,\ J_d=\frac{\kappa}{2}T_{H_1}^2(\delta c)_d,\ J_{D_2}=\frac{\kappa}{2}T_{H_1}^2(\delta c)_2,\\&J_l=\frac{\kappa}{2}T_M^2(\delta c)_l,\ J_r=\frac{\kappa}{2}T_N^2(\delta c)_r,
	\end{split}
\end{equation}
where 
\begin{equation}
	\begin{split}
		\kappa = \frac{\pi^2 k_{\text{B}}^2}{3h},\ P_{H_1}=P_{H_2}=\frac{e^2 V_0^2}{h} \frac{(\nu-\nu_i)\nu_i}{2\nu},
	\end{split}
\end{equation}
and $T_M, T_N, T_{H_1}, T_{H_2}$ are the temperatures at
$M, N, H_1, H_2$, respectively.
The dissipated powers $P_{H_1}, P_{H_2}$ arise due to the Joule heating of the associated voltage drop at $H_1, H_2$, respectively. We obtain $P_{H_1}=P_{H_2}$, since the voltage drop only depends on $\nu$ and $\nu_i$ \cite{PhysRevB.101.075308}. Here $(\delta c)_i = |(c_{\text{down}}-c_{\text{up}})_i|$ is the modulus of difference between the central charges of downstream and upstream modes in the $i$th segment 
of the deivce. 

Thereby we find 
\begin{equation}
	\begin{split}
		&\frac{\kappa}{2}T_{H_1}^2(\delta c)_2+\frac{\kappa}{2}T_{H_1}^2(\delta c)_d=P_{H_1}+\frac{\kappa}{2}T_M^2(\delta c)_l,\\&0+\frac{\kappa}{2}T_{H_1}^2(\delta c)_d=\frac{\kappa}{2}T_{N}^2(\delta c)_r,\\&\frac{\kappa}{2}T_{H_2}^2(\delta c)_1+\frac{\kappa}{2}T_{H_2}^2(\delta c)_u=P_{H_2}+\frac{\kappa}{2}T_N^2(\delta c)_r,\\&0+\frac{\kappa}{2}T_{H_2}^2(\delta c)_u=\frac{\kappa}{2}T_M^2(\delta c)_l,
	\end{split}
\end{equation}
and hence,
\begin{equation}
	\begin{split}
		&[(\delta c)_2 + (\delta c)_d]\frac{\kappa}{2}T_{H_1}^2=P_{H_1}+\frac{\kappa}{2}(\delta c)_lT_M^2,\\&[(\delta c)_1 + (\delta c)_u]\frac{\kappa}{2}T_{H_2}^2=P_{H_2}+\frac{\kappa}{2}(\delta c)_rT_N^2.
	\end{split}
\end{equation}
Therefore,
\begin{equation}
	\begin{split}
		&T_N^2[ \frac{(\delta c)_r}{(\delta c)_d}((\delta c)_2 + (\delta c)_d) + (\delta c)_r ] \\&= T_M^2[ \frac{(\delta c)_l}{(\delta c)_u}((\delta c)_1 + (\delta c)_u) + (\delta c)_l ].
	\end{split}
\end{equation}
We note that $(\delta c)_u=(\delta c)_d,\ (\delta c)_2=(\delta c)_1,\ (\delta c)_l=(\delta c)_r,\ (\delta c)_S=(\delta c)_G$, hence $T_M=T_N$. Thus we find
\begin{equation}\label{TcBBAB}
	\begin{split}
		&T_M=T_N\\&=\frac{eV_0}{\pi k_{\text{B}}}\sqrt{\frac{3(\nu-\nu_i)\nu_i}{\nu}}\Bigg[ \frac{(\delta c)_r}{(\delta c)_d}((\delta c)_2 + (\delta c)_d) - (\delta c)_l \Bigg]^{-\frac{1}{2}}.
	\end{split}
\end{equation}

\subsubsection{Noise spot contributions}
To calculate the contributions $\langle (\Delta I_S)^2 \rangle,\ \langle (\Delta I_G)^2 \rangle$
from the noise spots $O, P$, respectively,
we can use the same equations as in \cref{BABBIs}.

\section{Computation of CCC values}\label{CCCvalues}
Here we compute the CCC for specific choices of $\{\nu,\nu_i\}$ and for the Pf, APf, and PHPf states and different edge equilibration regimes.

\subsection{$\{\nu,\nu_i\}$ = $\{5/2,2\}$ with an APf state and full thermal equilibration}
Here heat transport is ballistic in ``outer", antiballistic in ``line", and ballistic in
``upper" segments, and we consult \cref{BABB} and \cref{heat}.
From \cref{TcBABB} with $\{\nu,\nu_i\}$ = $\{5/2,2\},\ (\delta c)_r=1/2,\ (\delta c)_1=3/2$, and $(\delta c)_d=2$ we find 
\begin{equation}
	\begin{split}
		T_M=T_N=\frac{e V_0}{\sqrt{5} \pi k_{\text{B}}}.
	\end{split}
\end{equation}	
	
Here we have $l_{\text{eq}}=l_{\text{eq}}^{\text{full}}$. We note that the ``outer" segment is in the ballistic regime. Hence, we expect an exponential suppression of $\langle (\Delta I_S)^2 \rangle = \langle (\Delta I_G)^2 \rangle$ \cite{PhysRevB.101.075308}.
Therefore, we find 
\begin{equation}
	\begin{split}
		\delta^2 I_1 =\delta^2 I_2 = |\delta^2 I_c| = \frac{8 e^3 V_0}{5 \sqrt{5} \pi h} \approx 0.227  \frac{e^3 V_0}{h}.
	\end{split}
\end{equation}
The source current is $I=5 e^2 V_0/(2 h)$, while the transmission coefficient is $t=2/(5/2)=4/5$. We thus obtain the corresponding Fano factors,
\begin{equation}
	\begin{split}
		F_1=F_2=|F_c|=\frac{\delta^2 I_1}{2 e I t(1-t)} = \frac{2}{\sqrt{5} \pi} \approx 0.284.
	\end{split}
\end{equation}

\subsection{$\{\nu,\nu_i\}$ = $\{5/2,7/3\}$ with a PHPf state and full thermal equilibration}
Here heat transport is ballistic in ``outer", antiballistic in ``line", and ballistic in
``upper" segments, and we consult \cref{BABB} and \cref{heat}.
From \cref{TcBABB} with $\{\nu,\nu_i\}$ = $\{5/2,7/3\},\ (\delta c)_r=1/2,\ (\delta c)_1=5/2$, and $(\delta c)_d=3$ we find 	
\begin{equation}
	\begin{split}
		T_M=T_N=\frac{\sqrt{7} e V_0}{15 \pi k_{\text{B}}}.
	\end{split}
\end{equation}	
	
Here we have $l_{\text{eq}}=l_{\text{eq}}^{\text{full}}$. We note that the ``outer" segment is in the ballistic regime. Hence, we expect an exponential suppression of $\langle (\Delta I_S)^2 \rangle = \langle (\Delta I_G)^2 \rangle$ \cite{PhysRevB.101.075308}.
Therefore, we find 
\begin{equation}
	\begin{split}
		\delta^2 I_1 =\delta^2 I_2 = |\delta^2 I_c| = \frac{28 \sqrt{7} e^3 V_0}{675 \pi h} \approx 0.0349  \frac{e^3 V_0}{h}.
	\end{split}
\end{equation}

The source current is $I=5 e^2 V_0/(2 h)$, while the transmission coefficient is $t=(7/3)/(5/2)=14/15$. We thus obtain the corresponding Fano factors,
\begin{equation}
	\begin{split}
		F_1=F_2=|F_c|=\frac{\delta^2 I_1}{2 e I t(1-t)} \approx 0.112.
	\end{split}
\end{equation}

\subsection{$\{\nu,\nu_i\}$ = $\{5/2,7/3\}$ with an APf state and full thermal equilibration}
Here heat transport is ballistic in ``outer", antiballistic in ``line", and ballistic in
``upper" segments, and we consult \cref{BABB} and \cref{heat}.
From \cref{TcBABB} with $\{\nu,\nu_i\}$ = $\{5/2,7/3\},\ (\delta c)_r=3/2,\ (\delta c)_1=3/2$, and $(\delta c)_d=3$ we find 	
\begin{equation}
	\begin{split}
		T_M=T_N=\frac{ \sqrt{7} e V_0}{\sqrt{45} \pi k_{\text{B}}}.
	\end{split}
\end{equation}	
	
Here we have $l_{\text{eq}}=l_{\text{eq}}^{\text{full}}$. We note that the ``outer" segment is in the ballistic regime. Hence, we expect an exponential suppression of $\langle (\Delta I_S)^2 \rangle = \langle (\Delta I_G)^2 \rangle$ \cite{PhysRevB.101.075308}.
Therefore, we find 
\begin{equation}
	\begin{split}
		\delta^2 I_1 =\delta^2 I_2 = |\delta^2 I_c| = \frac{28 \sqrt{7} e^3 V_0}{45 \sqrt{45} \pi h} \approx 0.078  \frac{e^3 V_0}{h}.
	\end{split}
\end{equation}
The source current is $I=5 e^2 V_0/(2 h)$, while the transmission coefficient is $t=(7/3)/(5/2)=14/15$. We thus obtain the corresponding Fano factors,
\begin{equation}
	\begin{split}
		F_1=F_2=|F_c|=\frac{\delta^2 I_1}{2 e I t(1-t)} \approx 0.2507.
	\end{split}
\end{equation}

\subsection{$\{\nu,\nu_i\}$ = $\{5/2,7/3\}$ with a PHPf state and partial thermal equilibration}
Here heat transport is ballistic in ``outer", antiballistic in ``line", and ballistic in
``upper" segments, and we consult \cref{BABB} and \cref{heat}.
We note that for partial thermal equilibration,
the counter-propagating modes in the lowest Landau levels of $\nu$ and $\nu_i$ localize in the ``line" segment and do not contribute to the CCC, while the modes in the second Landau levels do contribute. We effectively have the fillings $\{\nu^{\text{eff}}, \nu_i^{\text{eff}}\}=\{1/2,1/3\}$ for $\{\nu, \nu_i\}=\{5/2,7/3\}$. 
From \cref{TcBABB} with $\{\nu^{\text{eff}}, \nu_i^{\text{eff}}\}=\{1/2,1/3\},\ (\delta c)_r=1/2,\ (\delta c)_1=1/2$, and $(\delta c)_d=1$ we find 
\begin{equation}
	\begin{split}
		T_M=T_N=\frac{e V_0}{\sqrt{3} \pi k_{\text{B}}}.
	\end{split}
\end{equation}

Here we have $l_{\text{eq}}=l_{\text{eq}}^{\text{par}}$. We note that the ``outer" segment is in the ballistic regime. Hence, we expect an exponential suppression of $\langle (\Delta I_S)^2 \rangle = \langle (\Delta I_G)^2 \rangle$ \cite{PhysRevB.101.075308}.
Therefore, we find 
\begin{equation}
	\begin{split}
		\delta^2 I_1 =\delta^2 I_2 = |\delta^2 I_c| = \frac{4e^3 V_0}{9 \sqrt{3} \pi h} \approx 0.0816  \frac{e^3 V_0}{h}.
	\end{split}
\end{equation}
The source current is $I=5 e^2 V_0/(2 h)$, while the transmission coefficient is $t=(7/3)/(5/2)=14/15$. We thus obtain the corresponding Fano factors,
\begin{equation}
	\begin{split}
		F_1=F_2=|F_c|=\frac{\delta^2 I_1}{2 e I t(1-t)} \approx 0.2625.
	\end{split}
\end{equation}

\subsection{$\{\nu,\nu_i\}$ = $\{5/2,7/3\}$ with an APf state and partial thermal equilibration}
Here heat transport is antiballistic in ``outer", antiballistic in ``line", and ballistic in
``upper" segments, and we consult \cref{ABABB} and \cref{heatABABB}.
We note that for partial thermal equilibration,
the counter-propagating modes in the lowest Landau levels of $\nu$ and $\nu_i$ localize in the ``line" segment and do not contribute to the CCC, while the modes in the second Landau levels do contribute. We effectively have the filling $\{\nu^{\text{eff}}, \nu_i^{\text{eff}}\}=\{1/2,1/3\}$ for $\{\nu, \nu_i\}=\{5/2,7/3\}$. 
From \cref{TcABABB} with $\{\nu^{\text{eff}}, \nu_i^{\text{eff}}\}=\{1/2,1/3\},\ (\delta c)_G=1/2$ we find
\begin{equation}
	\begin{split}
		T_M=T_N=\frac{\sqrt{2}}{\sqrt{3}}\frac{eV_0}{\pi k_{\text{B}}}.
	\end{split}
\end{equation}

Here we have $l_{\text{eq}}=l_{\text{eq}}^{\text{par}}$. We note that the ``outer" segment is in the antiballistic regime. Therefore, a constant noise for $\langle (\Delta I_S)^2 \rangle = \langle (\Delta I_G)^2 \rangle$ is expected \cite{PhysRevB.101.075308}. We use Ref.~\onlinecite{PhysRevB.101.075308} to calculate $\langle (\Delta I_S)^2 \rangle = \langle (\Delta I_G)^2 \rangle$,
\begin{equation}\label{ABasym}
	\begin{split}
		\langle (\Delta I_S)^2 \rangle &= \langle (\Delta I_G)^2 \rangle = \frac{e^2}{h} \nu^{\text{eff}} \Bigg(\frac{\nu_{-}}{\nu_{+}}\Bigg) k_{\text{B}} T_M \\& \times \Bigg[ \frac{\sqrt{\pi}\Gamma(\frac{2+\alpha}{\alpha})}{2\Gamma(\frac{3}{2}+\frac{2}{\alpha})} + {}_2F_1\Bigg( -\frac{1}{2},\frac{2}{\alpha};\frac{2+\alpha}{\alpha};\frac{c_{\text{down}}}{c_{\text{up}}} \Bigg) \Bigg],
	\end{split}
\end{equation}
where $\nu^{\text{eff}} = 1/2, \nu_{+}=1, \nu_{-}=1/2, k_{\text{B}} T_M = \sqrt{2}eV_0/(\sqrt{3}\pi), \gamma=1, c_{\text{down}}=1, c_{\text{up}}=3/2, \alpha = -(c_{\text{down}} - c_{\text{up}}) \gamma \nu_{+}\nu_{-}/\nu^{\text{eff}} = 1/2$. We find $\langle (\Delta I_S)^2 \rangle = \langle (\Delta I_G)^2 \rangle \approx 0.07 e^3V_0/h$. 
Therefore, we obtain
\begin{equation}
	\begin{split}
		(\delta^2 I_1 =\delta^2 I_2) \approx 0.1543  \frac{e^3 V_0}{h},\ \delta^2 I_c \approx -0.0845 \frac{e^3 V_0}{h}.
	\end{split}
\end{equation}
The source current is $I=5 e^2 V_0/(2 h)$, while the transmission coefficient is $t=(7/3)/(5/2)=14/15$. We thus obtain the corresponding Fano factors,
\begin{equation}
	\begin{split}
		F_1=F_2=\frac{\delta^2 I_1}{2 e I t(1-t)} \approx 0.4959,\ |F_c| \approx 0.271.
	\end{split}
\end{equation}

\subsection{$\{\nu,\nu_i\}$ = $\{3,5/2\}$ with a Pf state and full thermal equilibration}
Here heat transport is ballistic in ``outer", antiballistic in ``line", and ballistic in
``upper" segments, and we consult \cref{BABB} and \cref{heat}.
From \cref{TcBABB} with $\{\nu,\nu_i\}$ = $\{3,5/2\},\ (\delta c)_r=1/2,\ (\delta c)_1=3$, and $(\delta c)_d=7/2$ we find 
\begin{equation}
	\begin{split}
		T_M=T_N\approx 0.077\frac{e V_0}{ k_{\text{B}}}.
	\end{split}
\end{equation}	
	
Here we have $l_{\text{eq}}=l_{\text{eq}}^{\text{full}}$. We note that the ``outer" segment is in the ballistic regime. Hence, we expect an exponential suppression of $\langle (\Delta I_S)^2 \rangle = \langle (\Delta I_G)^2 \rangle$ \cite{PhysRevB.101.075308}.
Therefore, we find 
\begin{equation}
	\begin{split}
		\delta^2 I_1 =\delta^2 I_2 = |\delta^2 I_c| \approx 0.128  \frac{e^3 V_0}{h}.
	\end{split}
\end{equation}

The source current is $I=3 e^2 V_0/h$, while the transmission coefficient is $t=(5/2)/3=5/6$. We thus obtain the corresponding Fano factors,
\begin{equation}
	\begin{split}
		F_1=F_2=|F_c|=\frac{\delta^2 I_1}{2 e I t(1-t)} \approx 0.155.
	\end{split}
\end{equation}

\subsection{$\{\nu,\nu_i\}$ = $\{3,5/2\}$ with a Pf state and partial thermal equilibration}
Here heat transport is ballistic in ``outer", antiballistic in ``line", and ballistic in
``upper" segments, and we consult \cref{BABB} and \cref{heat}.
We note that for partial thermal equilibration,
the counter-propagating modes in the lowest Landau levels of $\nu$ and $\nu_i$ localize in the ``line" segment and do not contribute to the CCC, while the modes in the second Landau levels do contribute. We effectively have the filling $\{\nu^{\text{eff}}, \nu_i^{\text{eff}}\}=\{1,1/2\}$ for $\{\nu, \nu_i\}=\{3,5/2\}$. 
From \cref{TcBABB} with $\{\nu^{\text{eff}}, \nu_i^{\text{eff}}\}=\{1,1/2\},\ (\delta c)_r=1/2,\ (\delta c)_1=1$, and $(\delta c)_d=3/2$ we find 
\begin{equation}
	\begin{split}
		T_M=T_N=0.5 \frac{e V_0}{\pi k_{\text{B}}}.
	\end{split}
\end{equation}

Here we have $l_{\text{eq}}=l_{\text{eq}}^{\text{par}}$. We note that the ``outer" segment is in the ballistic regime. Hence, we expect an exponential suppression of $\langle (\Delta I_S)^2 \rangle = \langle (\Delta I_G)^2 \rangle$ \cite{PhysRevB.101.075308}.
Therefore, we find 
\begin{equation}
	\begin{split}
		\delta^2 I_1 =\delta^2 I_2 = |\delta^2 I_c| = 0.5  \frac{e^3 V_0}{\pi h}.
	\end{split}
\end{equation}
The source current is $I=3e^2 V_0/(h)$, while the transmission coefficient is $t=(5/2)/3=5/6$. We thus obtain the corresponding Fano factors,
\begin{equation}
	\begin{split}
		F_1=F_2=|F_c|=\frac{\delta^2 I_1}{2 e I t(1-t)} \approx 0.19.
	\end{split}
\end{equation}

\subsection{$\{\nu,\nu_i\}$ = $\{3,5/2\}$ with an APf state and partial thermal equilibration}
Here heat transport is ballistic in ``outer", ballistic in ``line", and antiballistic in
``upper" segments, and we consult \cref{BBAB} and \cref{heatBBAB}.
We note that for partial thermal equilibration,
the counter-propagating modes in the lowest Landau levels of $\nu$ and $\nu_i$ localize in the ``line" segment and do not contribute to the CCC, while the modes in the second Landau levels do contribute. We effectively have the filling $\{\nu^{\text{eff}}, \nu_i^{\text{eff}}\}=\{1,1/2\}$ for $\{\nu, \nu_i\}=\{3,5/2\}$. 
From \cref{TcBBAB} with $\{\nu^{\text{eff}}, \nu_i^{\text{eff}}\}=\{1,1/2\},\ (\delta c)_r=3/2,\ (\delta c)_d=1/2, (\delta c)_2=1$, and $(\delta c)_l=3/2$ we find 
\begin{equation}
	\begin{split}
		T_M=T_N=0.5 \frac{e V_0}{\pi k_{\text{B}}}.
	\end{split}
\end{equation}

Here we have $l_{\text{eq}}=l_{\text{eq}}^{\text{par}}$. We note that the ``outer" segment is in the ballistic regime. Hence, we expect an exponential suppression of $\langle (\Delta I_S)^2 \rangle = \langle (\Delta I_G)^2 \rangle$ \cite{PhysRevB.101.075308}.
Therefore, we find 
\begin{equation}
	\begin{split}
		\delta^2 I_1 =\delta^2 I_2 = |\delta^2 I_c| = 0.5  \frac{e^3 V_0}{\pi h}.
	\end{split}
\end{equation}
The source current is $I=3e^2 V_0/(h)$, while the transmission coefficient is $t=(5/2)/3=5/6$. We thus obtain the corresponding Fano factors,
\begin{equation}
	\begin{split}
		F_1=F_2=|F_c|=\frac{\delta^2 I_1}{2 e I t(1-t)} = 0.6.
	\end{split}
\end{equation}

%